\documentclass[12pt]{article}
\usepackage{amsmath, amssymb}
\usepackage{graphicx}
\usepackage{bm}
\RequirePackage{tikz}
\usepackage{enumerate}
\usepackage[colorlinks,citecolor=blue]{hyperref}
\usepackage{color}
\usepackage{appendix}
\usepackage{mathrsfs}
\usepackage{float}
\usepackage{algorithm}
\usepackage{algpseudocode}
\usepackage{bigstrut}
\usepackage{dsfont}
\usepackage{cleveref}
\usepackage{subcaption}
\usepackage{booktabs}
\usepackage{pdflscape}
\usepackage{multirow}
\usepackage{natbib}
\usepackage{ntheorem}
\usepackage{authblk}
\usepackage{url} 
\usepackage{enumitem}
\usepackage[top=1in,bottom=1.3in,left=1in,right=1in]{geometry}
\usepackage{xr}

\graphicspath{{figures/}}

\newcommand{\E}{\mathbb{E}}
\newcommand{\K}[1]{\mathrm{K}\left(#1\right)}
\newcommand{\prob}{\mathbb{P}}

\newcommand{\tdomain}{\mathcal{X}}

\def\diff{{\rm d}}
\def\argmin#1{\mathrel{\mathop{\arg\min}\limits_{#1}}}

\newcommand{\bea}{\begin{eqnarray*}}
\newcommand{\eea}{\end{eqnarray*}}
\newcommand{\be}{\begin{eqnarray}}

\newcommand{\ee}{\end{eqnarray}}
\newcommand{\bsp}{\begin{split}}
\newcommand{\esp}{\end{split}}
\newcommand{\ed}{\end{document}}
\newcommand{\cat}{C}
\newcommand{\cps}{S}

\newcommand{\btab}{\begin{tabular}}

\newcommand{\etab}{\end{tabular}}
\newcommand{\bc}{\begin{center}}
\newcommand{\ec}{\end{center}}

\newcommand{\bi}{\begin{itemize}}
\newcommand{\ei}{\end{itemize}}
\newcommand{\bfi}{\begin{figure}}
\newcommand{\efi}{\end{figure}}
\newcommand{\ben}{\begin{enumerate}}
\newcommand{\een}{\end{enumerate}}
\newcommand{\bdes}{\begin{description}}
\newcommand{\edes}{\end{description}}

\newcommand{\bay}{\begin{array}}
\newcommand{\eay}{\end{array}}

\newcommand{\dtrain}{\mathcal{D}_{\mathrm{tra}} }
\newcommand{\dcal}{\mathcal{D}_{\mathrm{cal}} }
\newcommand{\RN}[1]{%
  \textup{\uppercase\expandafter{\romannumeral#1}}%
}

\def\bco{\iffalse}

\def\expit{\hbox{expit}}

\def\cp{\citep}

\def\F{Fr\'echet }

\def\M{\mathcal{M}}

\def\bco{\iffalse}

\def\reall{{\mathbb{R}}}

\def\expit{\hbox{expit}}

\def\cp{\citep}

\def\F{Fr\'echet }

\newcommand{\blind}{1}

\theoremstyle{plain}
\theoremheaderfont{\scshape}
\theoremseparator{.}
\theorembodyfont{\itshape}
\newtheorem{theorem}{\indent\large Theorem}
\newtheorem{prop}{\indent\large Proposition}
\newtheorem{corollary}{\indent\large Corollary}
\newtheorem{lemma}{\indent\large Lemma}

\makeatletter
\newtheoremstyle{proof}%
{\item[\hskip\labelsep\theorem@headerfont\MakeUppercase ##1\theorem@separator]}%
{\item[\hskip \labelsep\theorem@headerfont\MakeUppercase ##1\ ##3\theorem@separator]}
\makeatother

\theoremstyle{proof}
\theoremheaderfont{\scshape}
\theorembodyfont{\upshape}
\theoremsymbol{\ensuremath{\square}}

\theoremseparator{}
\newtheorem{assumption}{\indent}

\begin{document}

	\def\spacingset#1{\renewcommand{\baselinestretch}%
		{#1}\small\normalsize} \spacingset{1}

	
	\if1\blind
	{
		\title{\bf Conformal inference for random objects}
		\author[1]{Hang Zhou}
		\author[1]{Hans-Georg M\"{u}ller}
		\affil[1]{Department of Statistics, University of California, Davis}
		\maketitle
	} \fi

	\bigskip
	\begin{abstract}
		We develop an inferential toolkit for analyzing object-valued responses, which correspond to data situated in general metric spaces,  paired with Euclidean predictors within the conformal framework. To this end  we introduce  conditional profile average transport costs, where we compare distance profiles that correspond to one-dimensional distributions of probability mass falling into balls of increasing radius   through  the optimal transport cost when moving  from one distance profile to another.  The average transport cost  to transport a given distance profile to all others is crucial for   statistical inference in metric spaces and underpins the proposed conditional profile scores. A key feature of the proposed approach is to utilize  the  distribution of conditional profile average transport costs as conformity score  for general metric  space-valued  responses, which facilitates the construction of prediction sets  by  the split conformal algorithm. We derive the uniform convergence rate of the proposed conformity score estimators and establish asymptotic conditional validity for the prediction sets. The finite sample performance for synthetic data in various metric spaces demonstrates that the proposed conditional profile score outperforms existing methods in terms of  both coverage level and size of the resulting prediction sets,  even in the special case of scalar Euclidean responses. We also demonstrate the  practical utility of conditional profile scores for network data from New York taxi trips and for compositional data reflecting energy sourcing of U.S. states.
	\end{abstract}
	
	\noindent%
	\vfill
	
	\newpage
	\spacingset{1} 
	
\section{Introduction}\label{s:intr}

The conformal prediction framework was  introduced by \cite{vovk2005algorithmic, vovk2009line} as a sequential approach for forming prediction intervals. Subsequently, conformal inference has achieved notable success in  various statistical settings, such as predictive inference for non-parametric regression \citep{lei2013distribution, lei2014distribution, lei2018distribution, chernozhukov2021distributional,barber2022conformal},  covariate shift problems \citep{barber2022conformal, gibbs2021adaptive, gibbs2022conformal},   change point detection \citep{vovk2021retrain}, hypothesis testing \citep{vovk2021conformal,hu2023two}, outlier detection \citep{bates2023testing}, time series \citep{angelopoulos2023conformal, yang2024bellman} and survival analysis \citep{ConformalSurvivalAnalysis}.

In the regression setting with training data \((X_i, Y_i)_{i=1}^{n}\)  and additional  identically and independently distributed (i.i.d.) sampled data  \((X_{n+1}, Y_{n+1})\), the aim of conformal prediction is to construct a prediction set \(\hat{C}_{\alpha}(X_{n+1})\) such that
\begin{equation}\label{e:m}
	\prob({Y_{n+1}\in \hat{C}_{\alpha}(X_{n+1})})\geq 1-\alpha.
\end{equation}
To determine whether a value \(y\) in the response space should be included in the prediction set \(\hat{C}_{\alpha}(X_{n+1})\), the basic idea of conformal prediction is to test the null hypothesis that \(Y_{n+1}=y\) and to construct a valid \(p\)-value based on the empirical quantiles of a suitable score function  that is evaluated for the sample \((X_{1},Y_{1}),\ldots,(X_{n},Y_{n}), (X_{n+1},y)\).

Besides marginal coverage \eqref{e:m}, a more pertinent  but also more  ambitious and harder to achieve target is to require guaranteed coverage for each new instance rather than average coverage as
conveyed by \eqref{e:m}, i.e., to satisfy the conditional validity criterion
\begin{equation}\label{e:c}
	\prob({Y_{n+1}\in \hat{C}_{\alpha}(X_{n+1})\mid X_{n+1}})\geq 1-\alpha.
\end{equation}
The left-hand side of equation \eqref{e:c} represents coverage conditional on the predictor \(X_{n+1}\), while  
 the marginal coverage \eqref{e:m} is defined by taking an additional expectation over \(X_{n+1}\). In many real-world applications,  conditional validity is the more satisfactory criterion since often one aims at predictions  for a specific predictor level  \(X_{n+1}\), and averaging across all potential values of \(X_{n+1}\) provides a lesser guarantee  if one has \(X_{n+1}\)
 in hand and is interested in prediction at this specific value of the predictor.    However, conditional validity is hard to achieve and requires  strong assumptions for the distribution of \((X,Y)\) \citep{vovk2012conditional, lei2014distribution, foygel2021limits}. A commonly adopted  alternative is   asymptotic conditional validity 
 \citep{lei2018distribution, chernozhukov2021distributional}, i.e.,
\begin{equation}\label{e:d-cs}
	\prob({Y_{n+1}\in \hat{C}_{\alpha}(X_{n+1})\mid X_{n+1}})\geq 1-\alpha+o_{P}(1).
\end{equation}

A key feature of conformal inference is that the marginal  coverage level \eqref{e:m} is always guaranteed as long as the  score function meets certain symmetry conditions \citep{lei2018distribution}. However, the choice of the conformity score  influences the size of the prediction sets and a well chosen score yields smaller prediction sets. In particular, \cite{chernozhukov2021distributional} utilized an adjusted conditional distribution function as  conformity score and achieved an optimal prediction interval. However, their approach requires the optimization of a loss function involving the conditional quantile of \(Y\mid X\), which becomes rather  complicated when \(Y\mid X\) is not unimodal. It is also worth noting that the optimality in \cite{chernozhukov2021distributional} specifically concerns prediction sets that comprise a  single interval. In cases with bimodal conditional distributions, prediction sets featuring a union of distinct intervals are  expected to be more efficient than those featuring a  single interval. This  observation motivated the adoption by  \cite{izbicki2022cd} of the conditional distribution  as conformity score, demonstrating  that the resulting  HPD-split conformal prediction sets have the smallest Lebesgue measure asymptotically.


One method to achieve conditional coverage is to partition the sample space \(\mathcal{X}\) into distinct bins. For a new data point \(X_{n+1}\), the model is fitted and conformity scores are evaluated solely within the sub-region containing \(X_{n+1}\) \citep{lei2014distribution, izbicki2022cd}. These  approaches rely on the partitioning technique and specifically on the choice of tuning of parameters such as the number of bins. A general principle  is to  seek a conformity score that does not depend on $X$. 
The basic idea is straightforward: For any random variable \(X\) with a continuous distribution function \(F\), the transformed variable \(F(X)\) follows a uniform distribution on  \((0,1)\),  regardless  of \(X\). Building on this idea,  \cite{chernozhukov2021distributional} introduced the conditional cumulative distribution of \(Y\mid X\) as conformity score and \cite{izbicki2022cd} proposed  the conditional distribution of densities  as score function.

Extending the scope of previous models for conformal prediction, we  consider here a setting where the \(Y_i\) are complex data objects that are situated in a general metric space \(\mathcal{M}\) and the \(X_i\) are Euclidean predictors. Object data residing in metric spaces paired with Euclidean predictors have found increasing interest  in modern data analysis and various statistical approaches  
for analyzing such data have 
been developed over the last years.  Statistical models for regression scenarios with  object responses and Euclidean predictors 
have been  studied for various scenarios,  including  responses  located on a Riemannian manifold, which can be locally 
approximated by linear spaces \citep{chang1989spherical, fisher1993statistical, yuan2012local, thomas2013geodesic, cornea2017regression}, responses that are distributions located in the Wasserstein space \citep{chen2021wasserstein, zhu2021autoregressive} and also  for responses in general metric spaces \citep{petersen2019frechet, lin2021total}. These previous 
studies either implicitly or explicitly developed models for object regression through implementations of conditional Fr\'echet means, thus focusing on the mean response. 

For  real-world data analysis, understanding the distribution of the responses given a covariate level is as  important as quantifying  
the behavior of conditional means or \F means  when  covariates vary.  For example, regression models that only target conditional means 
are of little use when the underlying conditional distribution of a Euclidean response is not naturally centered around a single value, for example if it 
is bimodal.  We extend the  conformal framework to a new realm  by introducing  a conformity score that produces prediction sets of reasonably 
small size for all covariate levels,  is  sufficiently flexible to adapt to various  response distributions, is efficient and,  importantly, is easily computable  
for all types of responses that are situated in various non-Euclidean  metric spaces.

Mapping  object-valued data to  linear spaces such as tangent spaces for the case of random objects situated on Riemannian manifolds is a familiar  strategy
to circumvent 
 the absence of linear operations in metric spaces.  However, available transformations are limited to responses that are situated in distributional and Riemannian
  spaces and do not cover other metric spaces. Another  major limitation is that these linearizing maps  are either   metric-distorting or  
   not bijective. In the latter case inverse maps that are necessary  for the construction of prediction sets do not exist;  various ad hoc work-arounds have been 
   proposed, none of which is entirely 
 satisfactory \citep{petersen2016functional, bigo:17, chen2021wasserstein}. More recently, new methods that operate intrinsically and do not 
 rely on mapping to a linear space have been considered. These are more promising as they directly address the challenges of working within
 the non-Euclidean geometry of the response space \citep{dubey2020functional,zhu2021autoregressive}.  We adopt here such an intrinsic approach by adopting distance profiles  \citep{dubey2022depth} 
  for the proposed 
 conformal inference. 
Distance profiles characterize the distributions of the distances of each element to a 
random object in the metric space. Distance profiles are determined by both the metric of the object space and its underlying probability measure and they characterize this measure if the metric is of strong negative type. Distance profiles correspond to one-dimensional distributions  indexed by the elements $\omega\in\mathcal{M}$ of the metric space and 
form a well-defined stochastic process on the metric space. 
For  the proposed extension of conformal inference, we  introduce  \emph{conditional distance profiles}  \(F_{\omega,x}\), which  characterize the inherent conditional distribution  \(Y\mid X=x\). 

Statistical inference for object-valued data not only suffers from the absence of linear operations but also from the challenge that one does not have density functions, so that  the distribution function- and  density-based methods of \cite{chernozhukov2021distributional, izbicki2022cd} are not feasible anymore. Note that \(\{F_{\omega,x}:\omega\in\mathcal{M}\}\) is a family of one-dimensional distributions indexed by \(\omega \in \mathcal{M}\)  and 
 \(F_{Y,x}\) is a random measure when considering a random element $Y$ in the response space.  
Then the  expected value  of the 1-Wasserstein distance between \(F_{Y,x}\) and \(F_{\omega,x}\) characterizes the average transport cost of moving from \(F_{\omega,x}\) to \(F_{Y,x}\) and this motivates to employ \emph{conditional profile average transport costs}  to quantify  the compatibility of a given element 
\(\omega \in \mathcal{M}\) with  the conditional distribution of \(Y\mid X=x\). The heuristic is that less compatible elements should not be included in 
conditional prediction sets. Thus conditional profile costs serve as proxies for the unavailable conditional density function in general metric spaces and provide the starting point to arrive at conformal inference for  object data by employing  local linear estimators for both  conditional distance profiles  and  conditional profile average transport costs. We derive uniform convergence rates, providing a solid theoretical foundation 
and show that these rates attain the optimal one-dimensional kernel smoothing rate when the metric space where responses reside has a polynomial covering number.

Employing this approach for conformal prediction  leads to  model-free  statistical inference for  object-valued responses coupled with Euclidean predictors when using   a \emph{conditional profile score}  as  conformity score, which we introduce here and that is defined as  the distribution of conditional profile average transport costs. We then use the split conformal algorithm to derive prediction sets for object responses and show that these prediction sets lead to  asymptotic conditional validity under mild assumptions. Conditional validity is also demonstrated via numerical experiments with synthetic data for various metric spaces. Even for the special Euclidean case  where the  responses are scalars, the proposed method performs as well as or better compared to existing  conformal methods, including \cite{romano2019conformalized, sesia2020comparison, chernozhukov2021distributional, izbicki2022cd}, across all simulation scenarios.  
When dealing with multivariate predictors,  local linear smoothing   becomes increasingly problematic  due to the curse of dimensionality.  For  this case  we therefore replace local linear smoothing by  single index  Fr\'echet regression  \citep{bhattacharjee2023single}, where one first obtains an estimate of a direction parameter and then projects the multivariate predictor onto this direction to obtain a single index predictor.  Under mild assumptions on the consistency of the direction parameter the asymptotic conditional validity of the proposed conformity score  remains valid.

To summarize,  the main contributions of this paper are as follows.  First, we introduce conditional  distance profiles for random objects paired with Euclidean predictors. Second, we propose a novel conditional profile average transport cost and demonstrate its utility  for statistical inference in general metric spaces. Third, we introduce the conditional profiles score, which is a new conformity score for  object responses situated in general metric spaces and paired with Euclidean predictors. Fourth, we show that this score achieves asymptotic conditional coverage under mild assumptions. Fifth,  we develop a theoretical framework to establish  uniform convergence rates for the local linear estimator involving  function classes defined on metric spaces. Sixth, the efficiency of the conditional profile score is illustrated through comprehensive simulations and data applications across various metric spaces. Even  for the classical case of scalar responses in $\mathbb{R}$ the proposed conditional profile  score is as good or outperforms existing conformal methods in terms of both coverage levels and sizes of prediction sets. Data illustrations include   networks for  New York taxi trips and to compositional data reflecting energy usage by U.S. states as responses. 

The paper is organized as follows. In Section \ref{s:cdp}, we introduce conditional distance profiles and conditional profile average transport costs. The main methods are presented in Section \ref{s:mth}, including the split conformal method and a theorem that provides a general condition  for estimators of the transport costs so that estimators that satisfy it generate  conformity scores with guaranteed asymptotic conditional validity. In Section \ref{s:thy}, we obtain the uniform convergence rates  of local linear estimators and show that they achieve asymptotic conditional validity. The multivariate predictor case is discussed in Section \ref{s:mpd}. Numerical results for simulated data are presented in Section \ref{s:sim}, and data applications are provided in Section \ref{s:rda}. The proofs and additional results can be found in the Supplement.

\section{Conditional distance profiles}\label{s:cdp}

In what follows, for random sequences $A_{n}$ and $B_{n}$, we use \(A_{n} = O_{p}(B_{n})\) to denote \(\mathbb{P}(A_{n} \leq MB_{n}) \geq 1 - \epsilon\)  and \(A_{n} = o_{p}(B_{n})\) for  \(\mathbb{P}(A_{n} \geq \epsilon B_{n}) \rightarrow 0\) as \(n \rightarrow \infty\) for each \(\epsilon > 0\) and a positive constant \(M\). A non-random sequence \(a_{n}\) is said to be \(O(1)\) if it is bounded, and for each non-random sequence \(b_{n}\), \(b_{n} = O(a_{n})\) stands for \(b_{n} / a_{n} = O(1)\) and \(b_{n} = o(a_{n})\) stands for \(b_{n} / a_{n} \rightarrow 0\). The relation \(a_n \lesssim b_n\) indicates \(a \leq \text{const} \cdot b\) for large \(n\), and the relation \(\gtrsim\) is defined analogously. We write \(a_n \asymp b_n\) if \(a_n \lesssim b_n\) and \(b_n \lesssim a_n\). For \(a \in \mathbb{R}\), we use \(\lfloor a \rfloor\) to denote the largest integer smaller or equal to \(a\). We write  \(\mathcal{L}^{2}(\mathcal{D}) := \{f: \,\, \mathcal{D} \mapsto \mathbb{R}: \,\, \int_{\mathcal{D}} f^2(s)\diff s < \infty\}\) for  the space of square-integrable functions on \(\mathcal{D}\), \(\|f\|_2^2 := \int_{\mathcal{D}} f(s)^2 \diff s\), and \(\|f\|_{\infty} := \sup_{s \in \mathcal{D}} |f(s)|\).

Consider a random pair \(Z = (X, Y) \in \mathcal{X} \times \mathcal{M}\), where \(\mathcal{X}\) is a compact subset of \(\mathbb{R}^{d}\) and \(\mathcal{M}\) is a totally bounded, separable metric space with the associated distance function \(d(\cdot, \cdot)\). Given a probability space \((\mathcal{T}, \mathscr{T} , \mathbb{P})\), where \(\mathscr{T}\) represents the Borel \(\sigma\)-algebra on the domain \(\mathcal{T}\) and \(\mathbb{P}\) is a probability measure, the random pair \(Z\) can be described as a measurable mapping \(Z: \,\,  \mathcal{T} \to \mathbb{R}^d \times \mathcal{M}\). The joint law of \((X, Y)\) is represented by \(\mathbf{P}_Z\), such that \(\mathbf{P}_Z(A) = \mathbb{P}(\tau \in \mathcal{T}:\,\, Z(\tau) \in A)\) for any Borel measurable set \(A \subset \mathbb{R}^d \times \mathcal{M}\). We denote the marginal laws of \(X\) and \(Y\) as \(\mathbf{P}_X\) and \(\mathbf{P}_Y\), respectively. We also assume the conditional probability measure $\mathbf{P}_{Y\mid x}$  of \(Y\) given \(X=x\) exists, where  \(Y\) is the object response and \(X\) a  Euclidean predictor.

For any \(x \in \mathcal{X}\) and \(\omega \in \mathcal{M}\), let \(F_{\omega, x}\) represent the cumulative distribution function (CDF)  of the distance between \(\omega\) and the response \(Y\), conditional on \(X = x\) and with respect to \(\mathbf{P}_{Z}\). We refer to  the \(F_{\omega, x}\) as \emph{conditional distance profiles}, 
\begin{equation}\label{e:d-dp}
	F_{\omega, x}(t) = \mathbb{P}(d(\omega,Y) \leq t \mid  X=x ), \ \text{for all } t \in \mathbb{R}^{+}.
\end{equation}
These conditional distance profiles extend the previously introduced distance profiles \(F_{\omega}(t) = \mathbb{P}(d(\omega, Y) \leq t)\) \citep{dubey2022depth} and related concepts \cp{wang:23:3}. They represent the probability distribution of \(Y \mid X = x\) around an element \(\omega\). When more probability mass of \(Y \mid X = x\) is concentrated near \(\omega\), the corresponding distance profile \(F_{\omega, x}(t)\) will have relatively larger values near \(t = 0\), compared to distance profiles for elements \(\omega \in \M\) with less probability mass.  

Distance profiles \eqref{e:d-dp} are  defined for all \( \omega \in \mathcal{M} \). If we observe $Y$, it is a realization of the random element $Y=Y(\tau)$ for some  fixed element $Y(\tau)  \in \mathcal{M}$ and  \( \tau \in \mathcal{T} \). We   define $ F_{Y,x}=F_{Y(\tau),x}$ ,  i.e., 
$$ F_{Y,x}(t) = \mathbb{P}_{Y'}(d(Y, Y') \leq t \mid X' = x), $$
where \((X', Y') \) is an independent copy of \( (X,Y) \), and \( \mathbb{P}_{Y'} \) denotes the probability taken over the conditional distribution of \( Y' \mid X' \).

We may view \(F_{\omega, x}(t)\) and $F_{Y,x}(t)$ as  elements in the Wasserstein space of distributions with positive domain, 
\[ \mathcal{W}:=\,\left \{\mu \in \mathcal{P}(\mathbb{R}^{+}):\,\,\int_{\mathbb{R}^{+}}x^2 \diff\mu(x)<\infty \right \}, \]
where \(\mathcal{P}(\mathbb{R}^{+})\) is the set of all probability measures on \(\mathbb{R}^{+}\), equipped with the \(p\)-Wasserstein metric \(d_{W,p}(\cdot, \cdot)\), which for   \(\mu, \nu \in \mathcal{W}\) is defined as 
\begin{equation}\label{eq:def-kan}
	d_{W,p}(\mu,\nu):=\,\inf\left\{\left(\int_{\mathbb{R}^{+} \times \mathbb{R}^{+}}|x_{1}-x_{2}|^p \diff\Gamma(x_1,x_2)\right)^{1/p}: \,\, \Gamma \in \Gamma(\mu,\nu)\right\}\quad \text{for }p>0,
\end{equation}
where \(\Gamma(\mu, \nu)\) is the set of joint probability measures on \(\mathbb{R}^{+} \times \mathbb{R}^{+}\) with \(\mu\) and \(\nu\) as marginal measures. The Wasserstein space \((\mathcal{W}, d_{W,p})\) is a separable and complete metric space \citep{ambrosio2008, villani2009optimal}. The emerging field of distributional data analysis frequently utilizes  the Wasserstein metric for one-dimensional distributions 
\citep{petersen2016functional, pete:22, chen2021wasserstein}. We  write  \(F^{-1}(u) = \inf\{x \in \mathbb{R}:\,\,F(x) \geq u\}\) for \(u \in (0,1)\) to represent  quantile functions and 
consider both   \(F_{\omega, x}\), \(F^{-1}_{\omega,x}\) as representations of the  probability measure \(\mu_{\omega,x}\). 

The function \( F_{Y,x} \) indexed by \( Y \in \mathcal{M} \) can be regarded as a random element of \( \mathcal{W} \). For any \(\omega\in \mathcal{M}\), if \(F_{\omega, x}\) is absolutely continuous with respect to the Lebesgue measure, the optimal transport map \(F^{-1}_{Y,x} \circ F_{\omega, x}\) is  the push-forward map from \(F_{\omega, x}\) to \(F_{Y,x}\). The integral
\[d_{W,1}(F_{Y,x}, F_{\omega,x} )= \int_{\mathbb{R}^{+}}\left|F^{-1}_{Y,x}\circ F_{\omega, x}(u)-u \right| \diff F_{\omega, x}(u) \]
represents the 1-Wasserstein distance between \(F_{\omega, x}\) and \(F_{Y,x}\), and quantifies the amount of mass that needs to be {moved} from \(F_{\omega, x}\) to arrive at  \(F_{Y,x}\), i.e., the transport cost. 
 \begin{prop}[Proposition 2.17 of \cite{santambrogio2015optimal}]\label{p:ex}
	Given two cumulative distribution functions $F$ and $G$ defined on $\reall$,
	\begin{align*}
		\int_{0}^{1}\left|F^{-1}(u)- G^{-1}(u) \right|\diff u=\int_{\reall}\left|F(u)- G(u) \right|\diff u . 
	\end{align*}
\end{prop}
By Proposition \ref{p:ex}, the connection between the $d_{W,1}$-transport cost and the conditional distance profiles is as follows, 
$$d_{W,1}(F_{Y, x},F_{\omega, x} )= \int_{0}^{1}\left|F^{-1}_{Y, x}(u)- F^{-1}_{\omega, x}(u) \right|\diff u=
 \int_{0}^{\infty}\left|F_{Y, x}(u)- F_{\omega, x}(u) \right|\diff u,$$ and this motivates 
\begin{figure}[tbp]
    \centering
    \begin{subfigure}{0.32\textwidth}
        \centering
        \includegraphics[width=\linewidth]{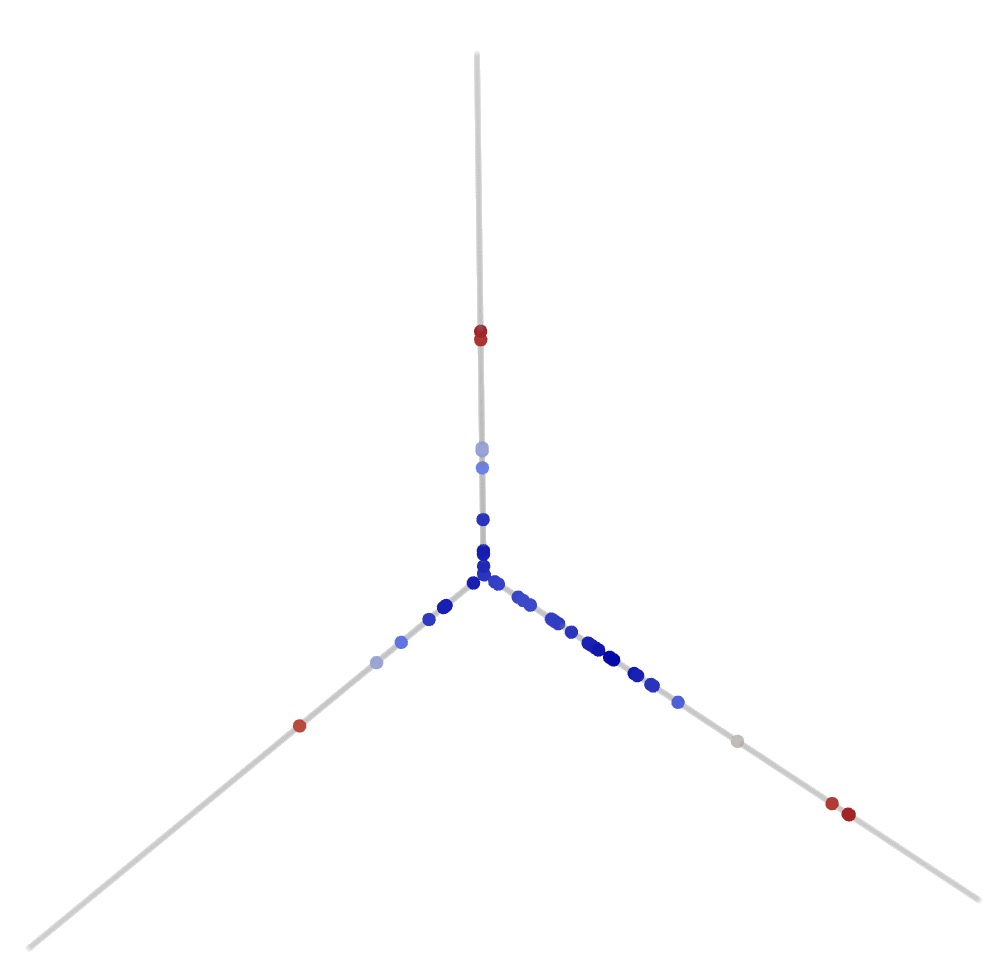}
    \end{subfigure}
    \hfill
    \begin{subfigure}{0.32\textwidth}
        \centering
        \includegraphics[width=\linewidth]{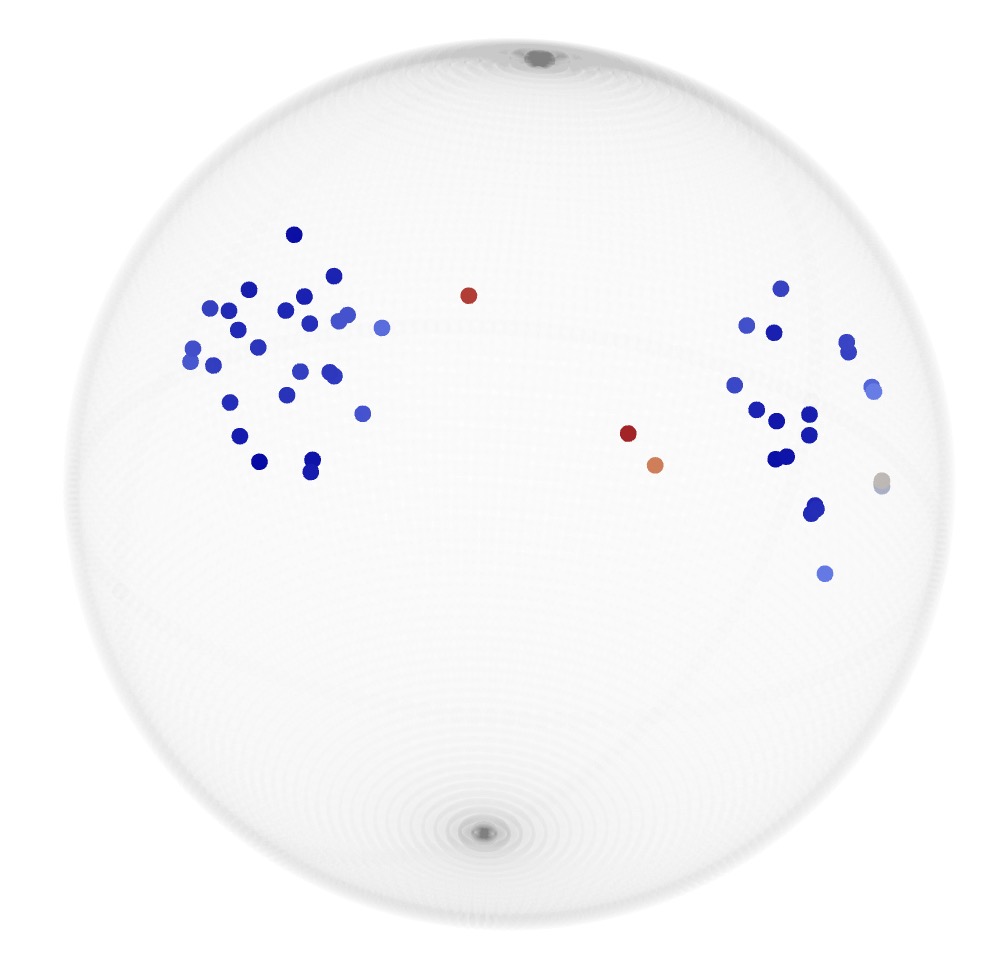}
    \end{subfigure}
    \hfill
    \begin{subfigure}{0.32\textwidth}
        \centering
        \includegraphics[width=\linewidth]{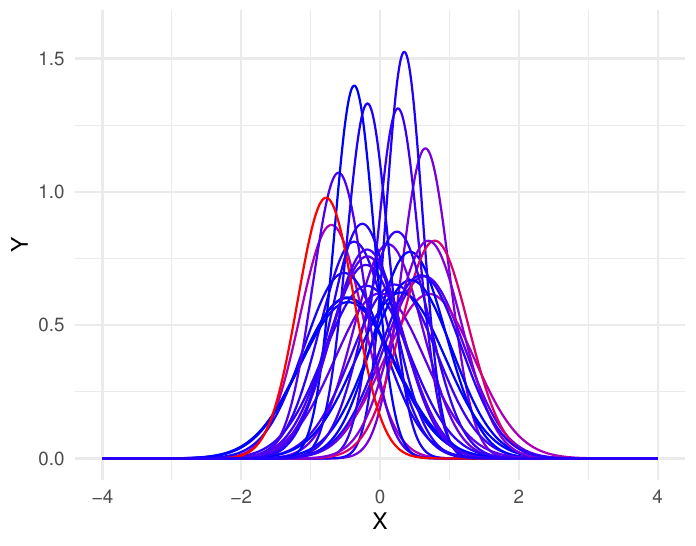}
    \end{subfigure}
    \caption{Illustration of the proposed profile  average transport costs for data points generated for  various metric spaces. The transition from low to high conditional transport cost  \eqref{e:d-cpc} is indicated by the color gradient from blue to red, where 
    red  indicates high values of profile average transport costs, while blue indicates low values. The left panel shows data  in the tree space \(\mathbb{T}^{3}\) with the BHV metric; each axis represents a distinct tree topology, and the position on the axis reflects the length of the interior edge. The middle panel shows data points that follow a distribution on \(\mathbb{S}^2\), which is characterized by two modes centered at  \(\mu_1=(1,0,0)\) and \(\mu_2=(0,1,0)\),
each with  equal probability  0.5. The data points are generated from the exponential map at \(\mu_k\),   applied to a random vector \(V_k\). 
    Here \(V_1=(0,\epsilon_1,\epsilon_2)\) and \(V_2=(\epsilon_3,0,\epsilon_4)\), where \(\epsilon_1,\dots,\epsilon_4\) are i.i.d. random variables drawn from \(\mathcal{N}(0,0.2)\). The right panel refers to data from the 2-Wasserstein space. Each curve represents the density function of a normal distribution  \(\mathcal{N}(\mu,\sigma)\), where  \(\mu\) and \(\sigma\) are drawn from uniform distributions \(\operatorname{Unif}(-0.8,0.8)\) and \(\operatorname{Unif}(0.25,0.75)\).
    \label{f:iX}}
\end{figure}
the concept of \emph{conditional profile average transport cost} (CPC),
\begin{equation}\label{e:d-cpc}
    C(\omega\mid x) =\,\mathbb{E} \left[\int_{\mathbb{R}^{+}} \left|F_{\omega, x}(t) - F_{Y, x}(t)\right| \, \mathrm{d}t  \, \Bigm|  X=x\right].
\end{equation}
The proposed cost function \( C(\omega \mid x) \) quantifies the average transport cost  moving from   \( F_{\omega,x} \) to \( F_{Y,x} \), where the expectation is taken over the conditional distribution of $Y\mid X=x$. A low value of \( C(\omega \mid x) \) indicates that this average transport cost is small, suggesting that the probability mass clusters more around \( \omega \) within the conditional distribution \( Y \mid X = x \). Figure \ref{f:iX} illustrates the  proposed CPCs, conditional on a fixed \(X=x_0\), for  three metric spaces: The tree space \(\mathbb{T}^{3}\) with the BHV metric \citep[\(\mathbb{T}^{3}\) denoting phylogenetic trees with three leaves and one interior edge, represented by the 3-spider formed by three rays identified at the origin]{billera2001geometry},  the sphere \(\mathbb{S}^2\) with the geodesic metric and the 2-Wasserstein space \(\mathcal{W}_2\) of distributions.  

\begin{figure}[tbp]
	\centering
	\includegraphics[width=0.7\textwidth]{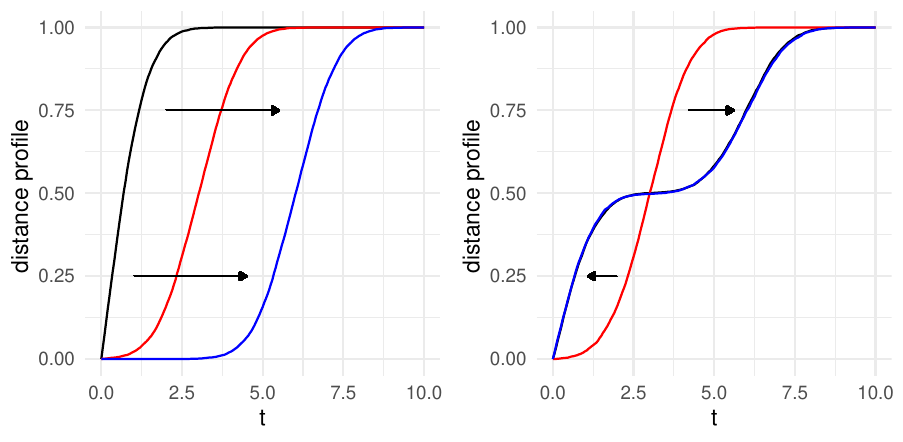}
	\caption{{Distance profiles for normal and mixture normal distributions for the special case where  \(\mathcal{M} = \mathbb{R}\). The underlying distribution is Gaussian \(\mathcal{N}(3,1)\) for the left panel and a mixture of Gaussians  \(0.5\mathcal{N}(-3,1) + 0.5\mathcal{N}(3,1)\) for the right panel. Black, red, and blue lines represent the estimated distance profiles at \(\omega = 3, 0, -3\), respectively. Arrows indicate the direction of transport of distance profiles, moving from the most central to the outermost  point as determined by \(\mathcal{H}(\omega\mid x_0)\). In the right panel, the black and blue lines overlap for the most part.}
	 \label{f:dis_pro}}
\end{figure}

We use a toy example to illustrate the difference between of proposed CPCs and the transport ranks defined in \cite{dubey2022depth}, which can also be extended to a conditional version that is given  by
\begin{equation}\label{e:dcm-c}
	\mathcal{H}(\omega\mid x)=\,\mathbb{E}\left[\int_0^1\left\{F_{Y, x}^{-1}(u)-F_{\omega, x}^{-1}(u)\right\} \mathrm{d} u\Bigm| X=x\right].
\end{equation}
Given any $X=x_{0}$, for a distribution $Y\mid X=x_{0}$ that is  symmetric around a single point \(\omega_{0}\), the integral in Equation \eqref{e:dcm-c} with \(\omega=\omega_{0}\) can be  expected to be relatively large and  \(\mathcal{H}(\omega\mid x_{0})\) to decrease as the distance \(d(\omega_{0},\omega)\) increases. The left panel of Figure \ref{f:dis_pro} displays estimators for the distance profiles \(F_{3, x_0}(t)\), \(F_{0, x_0}(t)\), and \(F_{-3, x_0}(t)\), corresponding to the conditional distribution \(Y\mid X=x_0\sim\mathcal{N}(3,1)\). In this scenario, \(\omega_{0}=3\) represents the most central point, with mass transferring from \(F_{\omega_0, x_0}(t)\) to \(F_{\omega, x_0}(t)\) from left to right for all \(t>0\) and \(\omega\neq \omega_{0}\). However, when the underlying distribution of \(Y\mid X=x_0\) is not centered around a single point, the most central point as determined by Equation \eqref{e:dcm-c} may not always be the most pertinent choice.  For instance, consider a mixture of normal distributions \(Y\mid X=x_0\sim0.5\mathcal{N}(-3,1)+0.5\mathcal{N}(3,1)\); the right panel of Figure \ref{f:dis_pro} illustrates estimators for distance profiles \(F_{3, x_0}(t)\), \(F_{0, x_0}(t)\), and \(F_{-3, x_0}(t)\). By symmetry, \(F_{3, x_0}(t) = F_{-3, x_0}(t)\) for all \(t>0\). Notably, mass transfers from \(F_{0, x_0}(t)\) to \(F_{3, x_0}(t)\) (or \(F_{-3, x_0}(t)\)) proceed from right to left for \(t\in(0,3)\) and from left to right for \(t\in(3,\infty)\). Since the transport rank  \(\mathcal{H}(\omega\mid x_0)\) reflects the difference between rightward and leftward moving mass, rather than the total transport cost, at the global center of the data, \(\omega_0=0\), positioned between the two modes, the integral in Equation \eqref{e:dcm-c} reaches its maximum and decreases as \(\omega\) moves away from \(\omega_0\). However, in bimodal settings, this global center is less pertinent and statistical inference that adapts to the mixture distribution is preferable.   The proposed  average transport cost criterion performs much better in bimodal cases, as illustrated in Figure \ref{f:yq} of Section \ref{s:mth} below. 
   
\section{Conformal inference for object data}\label{s:mth}
 
Given i.i.d. observations \((X_{i},Y_{i}) \in \mathcal{X} \times \mathcal{M}\) for \(i = 1, \cdots, n\), we aim to predict \(Y_{n+1}\) using the information from a future predictor \(X_{n+1}\). 
In contrast to standard   regression methods that correspond to versions of \F regression in the scenario with random object responses and focus on the conditional Fr\'echet mean, our goal is to construct a prediction set \(\hat{C}_{\alpha}(X_{n+1})\) that ensures asymptotic conditional validity \eqref{e:d-cs} for a specified coverage level \(1-\alpha\). The collection of prediction sets \(\hat{C}_{\alpha}:=\{\hat{C}_{\alpha}(x):\,\,x\in\mathcal{X}\}\) is referred to as the \emph{\(\alpha\)-level prediction set}. 
 
The selection of a good score function is crucial for the effectiveness of conformal inference methods. A well-chosen score not only yields a smaller prediction set but also achieves asymptotic conditional validity. Note that for any random variable \(X\) with continuous  distribution function \(F\), the transformed variable \(F(X)\) follows a uniform distribution on the interval \((0,1)\), regardless of \(X\). Building on this, \cite{chernozhukov2021distributional} proposed \(F(Y,X)\) as the conformity score, where \(F(y,x) = \mathbb{P}(Y \leq y \mid X = x)\) represents the conditional CDF of \(Y\) for a given \(X = x\). Similarly, \cite{izbicki2022cd} proposed the HPD-split score \(H(f(Y\mid X)\mid X)\), where \(f(y\mid x)\) is the conditional density function of \(Y\) given \(X = x\) and \(H(z\mid x) = \mathbb{P}(f(Y\mid X) \leq z \mid X = x)\) denotes the conditional CDF of \(f(Y\mid X)\) for a given \(X = x\).

%

\begin{figure}[tbp]
\centering
	\includegraphics[width=0.99\textwidth]{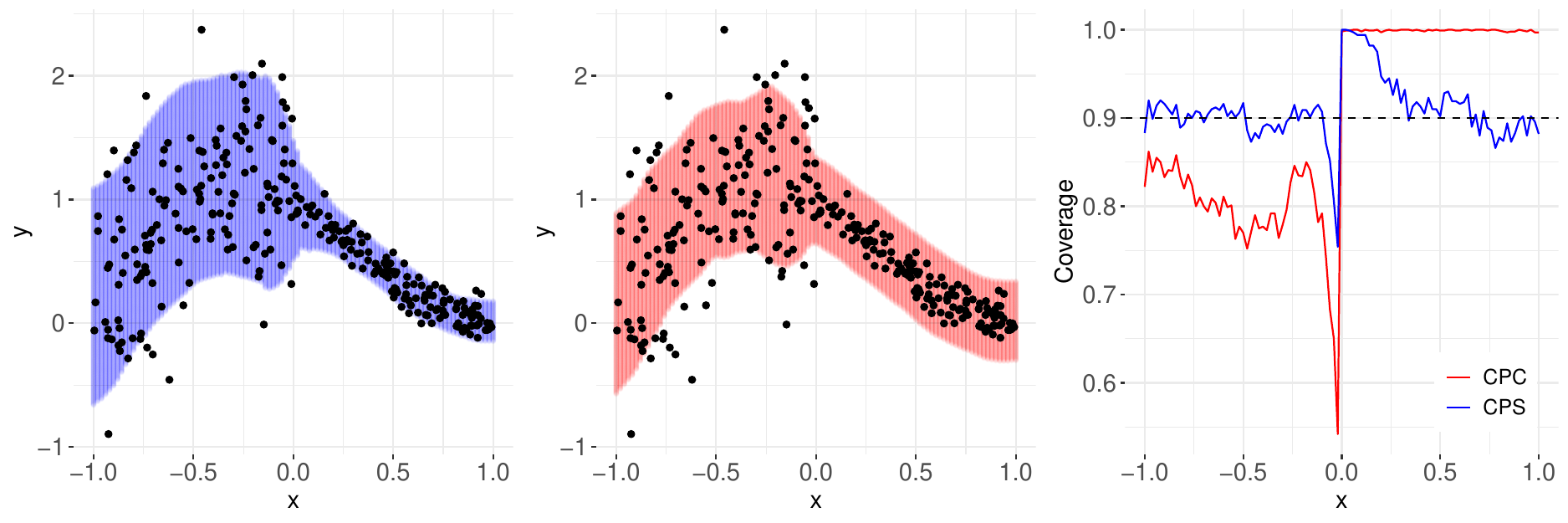}
    \centering
    \begin{subfigure}{0.32\textwidth}
        \centering
        \includegraphics[width=\linewidth]{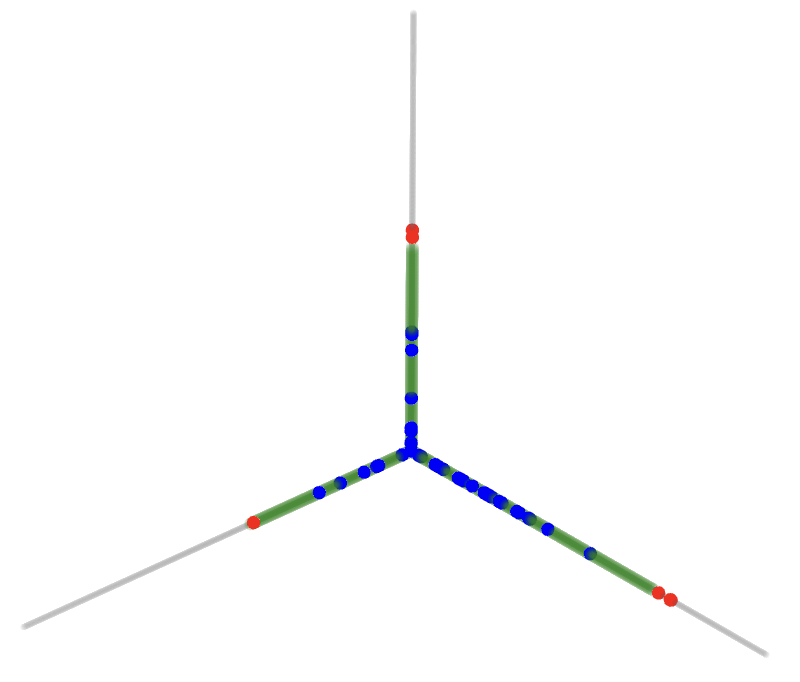}
    \end{subfigure}
    \hfill
    \begin{subfigure}{0.32\textwidth}
        \centering
        \includegraphics[width=\linewidth]{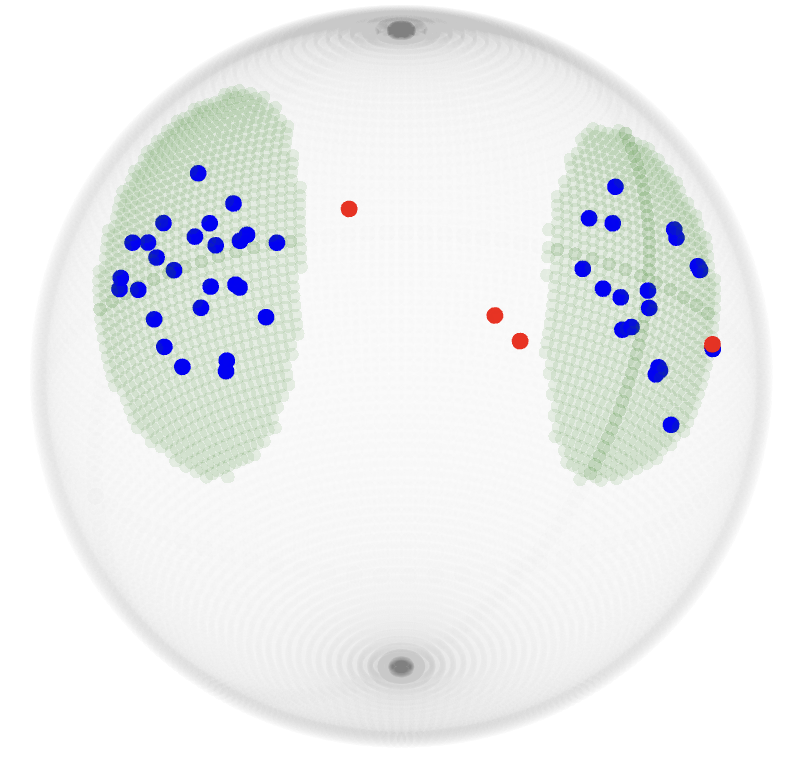}
    \end{subfigure}
    \hfill
    \begin{subfigure}{0.32\textwidth}
        \centering
        \includegraphics[width=\linewidth]{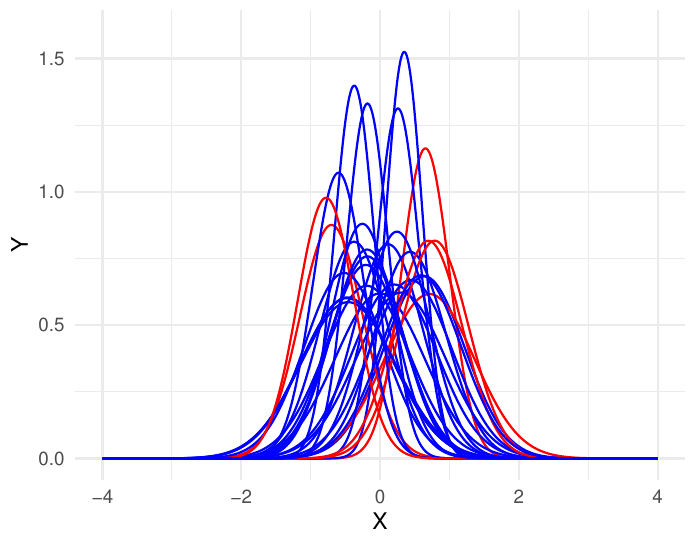}
    \end{subfigure}
    \caption{First row: Prediction sets obtained using conditional profile scores (CPS, left panel) and conditional profile costs (CPC, middle panel) as score functions with the split conformal algorithm. The data are generated from the model \( y_{i}=f(x_{i})+\sigma(x_{i})\epsilon_{i} \), where \( x_{i} \) are uniformly distributed on \((-1,1)\), \( f(x) = (x-1)^2(x+1) \), \( \sigma(x) = 0.5\mathds{1}_{\{x \leq 0\}} + 0.1\mathds{1}_{\{x > 0\}} \), and \( \epsilon_{i} \) are i.i.d.\,standard normal random variables. The target coverage level is $90\%$ and the  sample size is 2000, but only the first 200 data points are shown for illustration. The right panel displays the conditional coverage levels for both score functions (CPC in red and CPS in blue), evaluated on a test set. 
    Second row: Illustration of conformal sets using conditional profile scores based on the data from Figure \ref{f:iX}. The blue points in the left and middle panels and blue colored densities in the right panel represent the data falling within the 90\% prediction set, while the red points (curves) denote those that fall outside the 90\% prediction set. In the left and middle panels, the estimated  prediction sets are highlighted in green.
     \label{f:iC}}
\end{figure}

As discussed in Section \ref{s:cdp}, the conditional profile average transport costs $C(\omega\mid X=x)$ \eqref{e:d-cpc} measure the average transport mass from $F_{\omega, x}$ to $F_{Y, x}$ with respect to the conditional distribution $Y\mid X=x$. One direct approach is to use the CPCs \( C(Y_{i} \mid X_{i}) \) as the conformity score. However, this only ensures marginal validity because the distribution of the transport costs \( C(Y_{i} \mid X_{i}) \) can vary for different values of \( X_{i} = x \). Consequently, the level  \((1-\alpha)\) quantile of \( C(Y_{i} \mid X_{i}) \) derived from the calibration set is not a promising threshold across all covariates. This is illustrated in the first row of Figure \ref{f:iC}, which shows that in a heteroscedastic nonparametric regression setting, the coverage level of the prediction set based on the CPC score falls below the target for \( x \leq 0 \) and exceeds the target for \( x > 0 \).
To achieve conditional validity, we therefore introduce the \emph{conditional profile score} (CPS) $\cps( C(Y_{i} \mid X_{i})  \mid X_i)$ as a conformity score for random objects, where $\cps( \cdot\mid \cdot)$ is the conditional distribution of the profile-averaged transport costs:
\begin{equation}\label{e:d-pds}
	\cps(z\mid x) :=\, \mathbb{P}(\cat(Y\mid X) \leq z \mid  X=x).
\end{equation}

The split conformal method has become a popular tool due to its computational efficiency and the benefit of needing to train the model only once \citep{lei2018distribution, chernozhukov2021distributional, izbicki2022cd}. Its underlying principle is sample splitting, which ensures independence between the estimators and subsequent statistics. Sample splitting has a long history and it has been adopted for various problems beyond conformal inference,  including variable selection in high dimensions \citep{wasserman2009,meinshausen2009}, change point detection \citep{zou2020}, testing and false discovery rate control \citep{du2020}. 

Let \( \hat{F}_{\omega, x} \), \( \hat{C}(\omega\mid x) \), and \( \hat{S}(z\mid x) \) denote estimates of \( F_{\omega, x} \), \( C(\omega\mid x) \), and \( S(z\mid x) \), respectively. An outline of the algorithm for implementing the split conformal method with CPS is provided in Algorithm \ref{a:cf}. The second row of  Figure \ref{f:iC} illustrates the conformal sets derived by Algorithm \ref{a:cf} using the CPS \eqref{e:d-pds} as conformity scores for the data in Figure \ref{f:iX}, conditional on a specific \(X = x\). Note that the generated conformal prediction sets aptly provide conformal inference for both unimodal and bimodal structures across different metric spaces.

\begin{algorithm}[h]
\caption{Split conformal algorithm  for object valued data}
\label{a:cf}
\leftline{\textbf{Input:} Data $(X_{i},Y_{i})$, $i=1,\ldots,n$;  level $\alpha$ and a new data point  $X_{n+1}$.  }
\begin{algorithmic}[1] 
\State Randomly split $\{(X_{i},Y_{i})\}_{i=1}^{n}$ into training set $\dtrain$ and calibration set $\dcal$.
\State Get $\hat{F}_{\omega, x}(t),{\hat{\cat}} (\omega\mid x)$ and $\hat{\cps}(z\mid x)$ based on the training data $\dtrain$.
\State Evaluate the conformity scores $\{\hat{S}_{i}=\hat{\cps}(\hat{\cat}(Y_{i}\mid X_{i})\mid X_{i}) \} $ for $(X_{i},Y_{i})$ in the calibration set $\dcal$. 
\State   Compute $\hat{Q}_{\alpha} $, the $(1-\alpha)(1+1/|\dcal|)$ empirical quantile of $\{\hat{S}_{i}\}$.
\end{algorithmic}
\leftline{\textbf{Output:} Return the $(1-\alpha)$ prediction set $\hat{C}_{\alpha} (X_{n+1})=\{y\in\mathcal{M}:\,\, \hat{\cps}(\hat{\cat}(y\mid X_{n+1})\mid X_{n+1})\leq \hat{Q}_{\alpha}\}$.}
\end{algorithm}

When \( F_{\omega, x} \) is known for all \( \omega \in \mathcal{M} \) and \( x \in \mathcal{X} \), the conditional distribution of \( Y \mid X = x \) is fully determined if the metric space \( \mathcal{M} \) is of strong negative type \citep{dubey2022depth}, and thus $C(\omega\mid x)$ and $H(z\mid x)$ are known. Under the assumption that \( C(Y_{n+1} \mid X_{n+1}) \) has a continuous distribution function, \( S(C(Y_{n+1} \mid X_{n+1}) \mid X_{n+1}) \) follows a uniform distribution, which does not depend on the specific value of  \( X_{n+1} \). Thus, the population \( (1 - \alpha) \) quantile of \( S(C(Y_{n+1} \mid X_{n+1}) \mid X_{n+1}) \), denoted by \( Q_{\alpha} \), is \( 1 - \alpha \), and conditional validity is achieved since \( \mathbb{P}(S(C(Y_{n+1} \mid X_{n+1}) \mid X_{n+1}) \leq Q_{\alpha}) = 1 - \alpha \).
When  \( F_{\omega, x} \), \( C(\omega\mid x) \), and \( S(z\mid x) \) are unknown and need to be estimated,   the following structural condition on the convergence of the estimators that are utilized for this estimation step will  guarantee the asymptotic conditional validity of the resulting conformity score.

\begin{assumption}\label{as:a6}
	The conditional profile score $\cps(z\mid x)$ is Lipschitz continuous  in  both $z$ and $x $, that is, $\sup_{x\in\tdomain}| \cps(z_1\mid x)-\cps(z_2\mid x)|\leq L_{\cps}|z_{1}-z_{2}|$ and $\sup_{z\in\reall^+}|\cps(z\mid x_1)-\cps(z\mid x_2)|\leq L_{\cps}|x_{1}-x_{2}|$ for a positive constant $L_{\cps}$. 
\end{assumption}
\begin{assumption}\label{as:a7}
	For all \( n \in \mathbb{N}^{+} \) and i.i.d.\ \( (X_i, Y_i)_{i=1}^{n} \), the estimates $\hat{S}(z\mid x) $ and  $\hat{\cat}(\omega\mid x) $ satisfy  $\sum_{i=1}^{n} | \hat{\cps}(\hat{\cat}(Y_i \mid X_i) \mid X_i) - \cps(\cat(Y_i \mid X_i) \mid X_i) | = o_{P}(n) $. 
\end{assumption}
\begin{theorem}\label{thm:conf0}
	Under Assumptions  \ref{as:a6} and \ref{as:a7}, for the prediction set $\hat{C}_{\alpha}$ defined by Algorithm \ref{a:cf},
	$$\prob\left( Y_{n+1}\in \hat{C}_{\alpha}(X_{n+1}) \mid X_{n+1} \right)\geq 1-\alpha+o_{P}(1) .$$
\end{theorem}

The output of Algorithm \ref{a:cf}  is the prediction set \(\hat{C}_{\alpha}\), which  generally does not have an analytical form. Therefore, it is necessary to determine it over a finite grid  \(\mathcal{M}^{L}=\{y_{l}\}_{l=1}^{L}\) over  \(\mathcal{M}\). For example, if \(\mathcal{M}=\mathbb{S}^2\), one can first generate mesh grids \(\theta^{L}:=\{k\pi/L,\, k=1,2,\ldots,L\}\) and \(\phi^{L}:=\{2k\pi/L,\,  k=1,2,\ldots,L\}\). Then \(\mathcal{M}^{L^2}=\{(x,y,z):\,\, x=\sin(\theta_{l_1})\cos(\phi_{l_2}), y=\sin(\theta_{l_1})\sin(\phi_{l_2}), z=\cos(\theta_{l_1}), \, 1\leq l_1,l_2\leq L\}\).  The prediction sets then become \(\hat{C}_{\alpha} (X_{n+1})=\{y_{l}\in\mathcal{M}^{L^2}:\,\, \hat{\cps}(\hat{\cat}(y_l\mid X_{n+1})\mid X_{n+1})\leq \hat{Q}_{\alpha}\}\). The main computing cost is to obtain the  estimates \(\hat{F}_{\omega, x}\), \(\hat{\cat} (\omega\mid x)\), and \(\hat{\cps}(z\mid x)\). Thanks to the split conformal method,  one needs to compute these estimates only once.  With the score function estimates 
in hand, the evaluations of the scores of the  \(y_{i}\) are computationally  inexpensive.

\begin{figure}[tbp]
	\centering
	\includegraphics[width=0.7\textwidth]{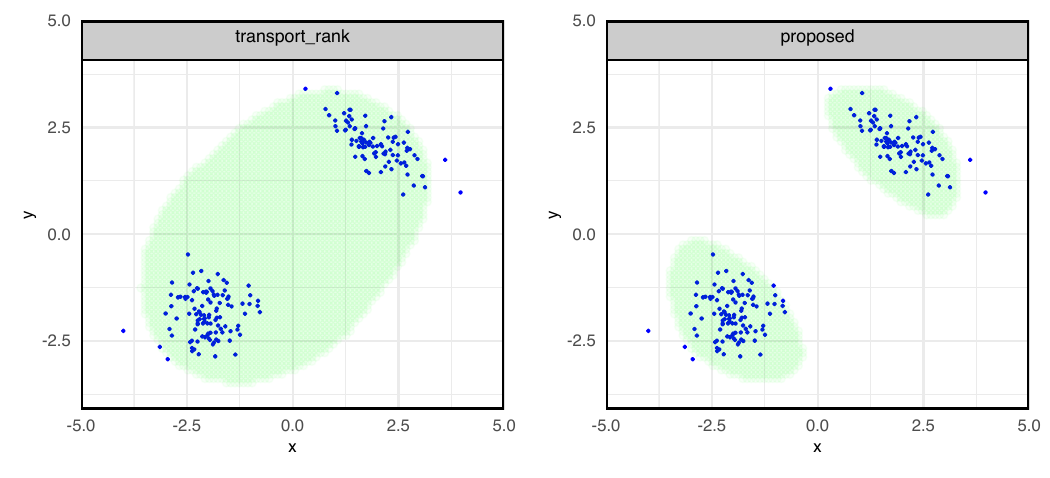}
	\caption{Conformal prediction sets generated using  transport ranks \citep{dubey2022depth} as conformity score (left panel) and the proposed conditional profile scores   defined in Equation \eqref{e:d-pds} (right panel), for   $\mathcal{M}=\mathbb{R}^2$. The training data are from a 2-dimensional Gaussian mixture, $ 0.5 \mathcal{N}(\boldsymbol{\mu}_{1},\boldsymbol{\Sigma}_1)+0.5 \mathcal{N}(\boldsymbol{\mu}_{2},\boldsymbol{\Sigma}_2)$, where $\boldsymbol{\mu}_{1}=(2,2)^{\top}$, $\boldsymbol{\mu}_{2}=(-2,-2)^{\top}$, $\bm{\Sigma}_1=(0.5,-0.3;-0.3,0.3)$ and $\bm{\Sigma}_2=(0.5,0;0,0.3)$. The data in the training set are  blue, and the respective {90\%} conformal sets  are shaded in green.
	 \label{f:yq}}
\end{figure}

An alternative is to use  conditional  transport ranks \citep{dubey2022depth} as conformity score. Based on \eqref{e:dcm-c}, unconditional transport ranks are obtained as 
\begin{equation}\label{eq:tr}
	R(\omega) = \expit\left( \E \left[ \int_{0}^{1}\{F^{-1}_{Y}(u) - F^{-1}_{\omega}(u) \}\,\diff u  \right] \right),
\end{equation}
where \(\expit(x) = \frac{e^{x}}{1 + e^{x}}\).  The transport ranks \(R(\omega)\) quantify the aggregated preference of \(\omega\) in relation to the data distribution, where a larger \(R(\omega)\) indicates that $\omega$ is more centrally located within the distribution.  However, as illustrated in Figure \ref{f:dis_pro}, \(R(\omega)\) is less suited to serve as a conformity score, which is evident  when  the underlying distribution is not centered around a single element. In the example in   Figure \ref{f:yq},  the conformal set determined by transport ranks as conformity scores is centered at the global center of the data and is seen to be suboptimal for a  2-dimensional mixture Gaussian distribution. In contrast, the proposed CPS successfully distinguishes the two groups and leads to smaller prediction sets.

\section{Estimation and theoretical results}\label{s:thy}

So far, conditional distance profiles, conditional profile average transport costs, and conditional profile scores have been introduced as population-level concepts. In subsection \ref{s:ll-e}, we focus on the case where \( \mathcal{X}\subset  \mathbb{R} \), employing local linear estimators for \( F_{\omega, x}(t) \), \( C(\omega \mid x) \), and \( S(z \mid x) \) based on independent random samples \(\{(X_i, Y_i)\}_{i=1}^{n}\) drawn from \((X, Y)\). These estimates are then combined with the split conformal algorithm to generate conformal prediction sets. The use of local linear estimates demonstrates that Assumption \ref{as:a7} and asymptotic conditional validity are achievable. Alternative estimation methods may also be employed, and provided they satisfy Conditions \ref{as:a6} and \ref{as:a7} in Theorem \ref{thm:conf0}, conditional validity is guaranteed.    In addition, we develop a theoretical framework to establish uniform convergence rates for the local linear estimator over function classes defined on metric spaces and derive optimal uniform convergence rates.

\subsection{Local linear estimates}\label{s:ll-e}

As we aim at  prediction sets for \(Y\) conditional on \(X=x\), it makes sense to primarily use those  \((X_{i},Y_{i})\)  for which \(X_{i}\) is close to \(x\) when aiming at  conditional estimates. This  motivates the adoption of   local linear smoothers \citep{fan1992variable, fan1993local} to obtain conditional empirical distance profiles and estimates of \(F_{\omega, x}(t)\) for each \(x\in\mathcal{X}\), \(\omega\in\mathcal{M}\), and \(t\in\mathbb{R}^{+}\), 
\begin{equation}\label{e:ll-f}
	\hat{F}_{\omega, x}(t)=\,\,\argmin{\beta_{0}\in\reall}\frac{1}{nh_n}\sum_{j=1}^{n}\left\{ L_{j}(\omega,t)-\beta_{0}-\beta_{1}(X_{j}-x) \right\}^2\K{\frac{X_j-x}{h_n}},
\end{equation}
where \(L_{j}(\omega,t)=\mathds{1}_{\{d(\omega,Y_{j})\leq t\}}\), \(\mathrm{K}(\cdot)\) is a symmetric and continuous density kernel on \([-1,1]\)  of bounded variation and \(h_{n}\) is a  sequence of bandwidths. Subsequently, to estimate the conditional profile average transport costs as defined in \eqref{e:d-cpc}, we utilize local linear smoothing for $(X_j,J_j(\omega,x))$, where \(J_j(\omega,x)= \int|\hat{F}_{\omega, x}(t)-\hat{F}_{Y_{j}, X_j}(t) |\mathrm{d} t\) with  estimated distance profiles \(\hat{F}_{\omega, x}\),
\begin{equation}\label{e:ll-l}
	\hat{\cat}(\omega\mid x)=\,\,\argmin{\beta_{0}\in\reall}\frac{1}{nh_n}\sum_{j=1}^{n}\left\{ J_{j}(\omega,x)-\beta_{0}-\beta_{1}(X_{j}-x) \right\}^2\K{\frac{X_j-x}{h_n}}.
\end{equation}
The estimated values of $S(z \mid x)$ are then
\begin{equation}\label{e:ll-h}
	\hat{\cps}(z\mid x)=\,\,\argmin{\beta_{0}\in\reall}\frac{1}{nh_n}\sum_{j=1}^{n}\left\{ H_{j}(z)-\beta_{0}-\beta_{1}(X_{j}-x) \right\}^2\K{\frac{X_j-x}{h_n}},
\end{equation}
where \(H_{j}(z)=\mathds{1}_{\{\hat{\cat}(Y_{j}\mid X_{j})\leq z \}}\) is the empirical estimate of \(\mathbb{P}(\cat(Y\mid X)<z)\).

\subsection{Theoretical results}\label{s:ll-s}

To obtain convergence rates of the estimates \(\hat{F}_{\omega, x}(t)\), \(\hat{\cat}(\omega\mid x)\), and \(\hat{\cps}(z\mid x)\) to their population targets,  
a key result is  the uniform convergence  of the following process, which is indexed by \(f \in \mathcal{F}\) and \(x \in \mathcal{X}\) $\, $ \citep{fan1992variable,fan1993local,hall1997role,choi1998bias}:
\begin{equation}\label{eq:def-An}
	A_{n,r}(x,f)=\sum_{j=1}^{n}f(X_{j},Y_{j})\K{\frac{X_{j}-x}{h_{n}}}(X_{j}-x)^r,\quad r=0,1,2,
\end{equation}
where $\mathcal{F}$ is a  generic  class of functions from $\tdomain \times \mathcal{M}$ to $\reall$. Let 
\begin{equation}\label{e:F1}
	\mathcal{F}_{1}=\,\{\mathds{1}_{\{d(\omega,y)\leq t \}}:\omega\in\mathcal{M},\ t\in \reall^{+} \}
\end{equation}
be the class of indicator functions indexed by $\omega$ and $t$. By considering $f_{0}(x,y)=1$ for all $x \in \tdomain$ and $y \in \mathcal{M}$, and $f_{\omega,t}(x,y)=\mathds{1}_{\{d(\omega,y)\leq t \}} \in \mathcal{F}_1$ for every $x \in \tdomain$, $\hat{F}_{\omega, x}$ as defined in \eqref{e:ll-f} has the  form:
$$
\hat{F}_{\omega, x}(t) = \,\frac{A_{n,2}(x,f_{0})A_{n,0}(x,f_{\omega,t})-A_{n,1}(x,f_{0})A_{n,1}(x,f_{\omega,t}) }{A_{n,2}(x,f_{0})A_{n,0}(x,f_{0})-A^2_{n,1}(x,f_{0})}.
$$
{Analogous expressions for \(\hat{\cat}(\omega\mid x)\) and \(\hat{\cps}(z\mid x)\) can be obtained by considering appropriate function classes \(\mathcal{F}\) in Equation \eqref{eq:def-An}.}

To derive the convergence rate and establish the asymptotic properties of \(A_{n,r}(x,f)\), 
we require   the following regularity assumptions.

\begin{assumption}\label{as:a1}
	The marginal distribution of $X$ has a continuous density function $f_X$, which satisfies $\inf_{x\in\operatorname{Support}(f_{X})} f_X(x)>c_1$ and $\sup_{x\in\tdomain} f_X(x)<c_{2}$ for  strict positive constants $c_1$ and $c_2$.
\end{assumption}
\begin{assumption}\label{as:a2}
	The bandwidth sequence $\{h_{n}\}_{n\geq 1}$ satisfies $nh_{n}/\log n\rightarrow\infty$  and $|\log h_n|/\log\log n\rightarrow\infty$ as $n\rightarrow\infty$.
\end{assumption}
\begin{assumption}\label{as:a3}
	The function class $\mathcal{F}$ is  bounded, i.e., there exists a  $M_\mathcal{F}>0$ such that  $$\sup_{f\in\mathcal{F}}\sup_{y\in\mathcal{M}}\sup_{x\in\tdomain} |f(x,y)|\leq M_{\mathcal{F}}<\infty.$$
\end{assumption}
Assumption \ref{as:a1}  is a mild condition widely adopted in kernel smoothing, while assumption \ref{as:a2} relates to a basic requirement for the bandwidth \(h_n\) that is necessary for consistency. Assumption \ref{as:a3} imposes a boundedness constraint on the function class \(\mathcal{F}\) that is satisfied by the function classes that we consider later.   Write  \(N(\epsilon, \mathcal{F}, d)\) for the minimal number of balls \(\{g: \,\, d(g, f) < \epsilon\}\) with radius \(\epsilon\) needed to cover \(\mathcal{F}\). 
For a function class \(\mathcal{F}\) that contains functions mapping from \(\mathcal{X}\times\mathcal{M}\) to \(\mathbb{R}\) and  has a finite-valued envelope function \(F_{\text{e}}\), we define the uniform covering number \(\mathcal{N}(\epsilon, \mathcal{F})\) of $\mathcal{F}$ as
$
\mathcal{N}(\epsilon, \mathcal{F}) :=\,\, \sup_{\mathbb{Q}} N(\epsilon\sqrt{\mathbb{E}_{\mathbb{Q}}[F_{\text{e}}^2]}, \mathcal{F}, d_{\mathbb{Q}}),
$
where the supremum is taken over all probability measures \(\mathbb{Q}\) on \(\mathcal{X}\times\mathcal{M}\) such that \(0 < \mathbb{E}_{\mathbb{Q}}[F_{\text{e}}^2] < \infty\). Here \(d_{\mathbb{Q}}\) is the \(\mathcal{L}_{\mathbb{Q}}^2\) metric, where for  any two functions \(f, g \in \mathcal{F}\), 
$
d_{\mathbb{Q}}^2(f,g) = \int\{f(x)-g(x)\}^2 \, \text{d}\mathbb{Q}(x).
$

The following lemma  establishes the uniform convergence rate for the process \(A_{n,r}(x,f)\).

\begin{lemma}\label{thm:An}
    Under Assumptions \ref{as:a1} to \ref{as:a3}, with probability $1$, there exists an absolute constant $C_1$, such that for $r=0,1,2$, 
	\begin{itemize}
		\item[a).] If $\mathcal{N}(\epsilon,\mathcal{F})\lesssim  \epsilon^{-v}$ for a constant $v>0,$
		\begin{equation}\label{eq:thm-An-1}
\lim _{n \rightarrow \infty} \frac{\sup _{f \in \mathcal{F}} \sup _{x \in \tdomain}\left|A_{n,r}(x, f)-\E A_{n,r}(x,f)\right|}{\sqrt{2 n h_n\left|\log h_n\right|}}\leq C_{1}.
	\end{equation}
	\item[b).] If $\log \mathcal{N}(\epsilon,\mathcal{F})\lesssim  \epsilon^{-v}$ for a constant $0< v<2$,
	\begin{equation}\label{eq:thm-An-2}
\lim _{n \rightarrow \infty} \frac{\sup _{f \in \mathcal{F}} \sup _{x \in \tdomain}\left|A_{n,r}(x, f)-\E A_{n,r}(x,f)\right|}{\sqrt{2 n h_n^{1-v/2} }}\leq C_{1} .
	\end{equation}
	\end{itemize}
\end{lemma}

{Lemma \ref{thm:An} establishes the uniform convergence rate of the process \(A_{n,r}\). It is the key tool  for obtaining uniform convergence rates for the local linear estimator with object data; as for all other results, the proof is in the Supplement.} The uniform covering number \(\mathcal{N}(\epsilon,\mathcal{F})\) characterizes the complexity of the function class \(\mathcal{F}\). When \(\mathcal{F}\) has a polynomial uniform covering number, equation \eqref{eq:thm-An-1} indicates that \(A_{n,r}\) typically achieves a one-dimensional non-parametric smoothing  rate. However, for a relatively complex \(\mathcal{F}\) where \(\log \mathcal{N}(\epsilon,\mathcal{F}) \lesssim \epsilon^{-v}\) for a constant \(0 < v < 2\), the process \(A_{n,r}\) has a slower uniform convergence rate. Our primary focus is on the function class
\(
\mathcal{F}_{1}=\{\mathds{1}_{\{d(\omega,y) \leq t\}}:\,\, \omega \in \mathcal{M}, t \in \mathbb{R}^{+}\}.
\)
Applying Lemma \ref{thm:An} with \(\mathcal{F}_{1}\), we obtain the uniform convergence rate for conditional distance profiles. To proceed, we require additional assumptions on the continuity of \(F_{\omega, x}(t)\). The following Assumption \ref{as:a4} requires that the distance profiles \(F_{\omega, x}(t)\) are continuous in \(t\) and have bounded density functions, and  Assumption \ref{as:a5} stipulates   that \(F_{\omega, x}(t)\) is Lipschitz continuous in both \(x\) and \(\omega\).


\begin{assumption}\label{as:a4}
	For every $\omega \in \mathcal{M}$ and $x\in \tdomain$,  the distance  profile $F_{\omega, x}$ is absolutely continuous with continuous density $f_{\omega, x}$ and there exist strict positive constants $c_{3}$ and $c_{4}$ such that $\inf _{t \in \operatorname{support}\left(f_{\omega, x}\right)} f_{\omega, x}(t)\geq c_3 $ and $\sup _{t \in \mathbb{R}^{+}} f_{\omega, x}(t)\leq c_{4}<\infty$. 
\end{assumption}
\begin{assumption}\label{as:a5}
	For every $\omega \in \mathcal{M}$ and $t\in \reall^{+}$,  $F_{\omega, x}(t)$ is second order differentiable and has bounded second order derivatives with respect to $x$. Moreover, there exists a constant $L'$ such that $|F_{\omega_{1}, x}(t)- F_{\omega_{2}, x}(t)|\leq L'd(\omega_{1},\omega_{2})$ for all $x\in\tdomain,t\in\reall^{+}$ and $\omega_{1},\omega_{2}\in \mathcal{M}$.
\end{assumption}

\begin{theorem}\label{thm:ll-F}
	Under Assumptions \ref{as:a1} - \ref{as:a5}, for the distance profile estimator $\hat{F}_{\omega, x}$ defined by \eqref{e:ll-f},
	\begin{itemize}
		\item[a).] If $\mathcal{N}(\epsilon,\mathcal{F}_1)\lesssim  \epsilon^{-v}$ for a constant $v>0,$ 
		$$\sup_{\omega\in\mathcal{M} }\sup_{x\in\tdomain}\sup_{t\in\reall^{+}}\left|\hat{F}_{\omega, x}(t)- F_{\omega, x}(t) \right|=O\left(\sqrt{\frac{|\log h_n |+\log n}{n h_{n}}} +h_n^2 \right) \text{ a.s.}.$$
		\item[b).] If $\log \mathcal{N}(\epsilon,\mathcal{F}_1)\lesssim  \epsilon^{-v}$ for a constant $0< v<2$,
		$$\sup_{\omega\in\mathcal{M} }\sup_{x\in\tdomain}\sup_{t\in\reall^{+}}\left|\hat{F}_{\omega, x}(t)- F_{\omega, x}(t) \right|=O\left(\sqrt{\frac{1}{n h_{n}^{1+v/2}}} +h_n^2 \right) \text{ a.s.}.$$
	\end{itemize}
\end{theorem}

It is important to note that the convergence rates in Theorem \ref{thm:ll-F} are uniform not just over \(x \in \mathcal{X}\), but also over  \(\omega \in \mathcal{M}\) and  \(t > 0\). When choosing an asymptotically  optimal bandwidth sequence to balance the bias and stochastic error terms and  if \(\mathcal{F}_{1}\) has a polynomial uniform covering number, Corollary \ref{c:F} below implies that 
\(\hat{F}_{\omega, x}\) converges to \(F_{\omega, x}\) at a typical  one-dimensional kernel smoothing rate. For each \(x\) and \(\omega\), the empirical estimates of the distributions corresponding to  distance profiles in the unconditional case  can be estimated 
at a parametric rate \citep{dubey2022depth}. However, when applying the kernel smoother to the predictor space $\mathcal{X}$, as needed to obtain conditional distance profiles, achieving a root-\(n\) rate using data falling into a local window is not feasible \citep{hall1999methods}. The achievable rate for the conditional case is as follows.  

\begin{corollary}\label{c:F}
	Under Assumptions \ref{as:a1} - \ref{as:a5}, if  $\mathcal{N}(\epsilon,\mathcal{F}_1)\lesssim  \epsilon^{-v}$ for a constant $v>0$ and $h_{n}\asymp(n/\log n)^{-1/5}$
		$$\sup_{\omega\in\mathcal{M} }\sup_{x\in\tdomain}\sup_{t\in\reall^{+}}\left|\hat{F}_{\omega, x}(t)- F_{\omega, x}(t) \right|=O\left(\left( \frac{n}{\log n}\right) ^{-\frac{2}{5}} \right) \text{ a.s.}.$$
\end{corollary}

For a complex metric space \(\mathcal{M}\) where \(\log \mathcal{N}(\epsilon,\mathcal{F}_1) \lesssim \epsilon^{-v}\) with \(0 < v < 2\), the uniform convergence rate of \(\hat{F}_{\omega, x}\) becomes
\[
\sup_{\omega\in\mathcal{M}}\sup_{x\in\mathcal{X}}\sup_{t\in\mathbb{R}^+}\left|\hat{F}_{\omega, x}(t)- F_{\omega, x}(t) \right|=O\left( n^{-\frac{4}{10+v}} \right),
\]
which falls within the range \((n^{-2/5},n^{-1/3})\). This rate is slower than the one-dimensional non-parametric smoothing  rate but faster than the two-dimensional non-parametric rate.

The uniform covering number of the function class \(\mathcal{F}_{1}\), containing solely indicator functions,  is determined  by the geometric properties of the object space \(\mathcal{M}\). The following result provides  a sufficient condition for \(\mathcal{F}_{1}\) to be a VC-subgraph class with polynomial uniform covering number, leveraging the geometric structure of \(\mathcal{M}\) and the properties of the indicator functions within \(\mathcal{F}_{1}\). A more detailed description of  VC (Vapnik--Chervonenkis) dimension and VC class can be found in the Supplement S1.

\begin{lemma}\label{lem:VCsb}
	Let $\mathcal{F}_1=\{\mathds{1}_{\{d(\omega,y)\leq t \}} \}$ be the function class indexed by $\omega\in \mathcal{M}$ and $t\in\reall^{+}$. If $\{y:d(\omega,y)\leq t,\, \omega\in\mathcal{M},\, t\in\reall^{+} \}$ forms a VC-class in $\mathcal{M}$, then  $\mathcal{F}_1$ is a VC-subgraph class.
\end{lemma}

Many commonly used metric spaces fulfill the condition stated in Lemma \ref{lem:VCsb}. This includes the Euclidean space and the sphere \(\mathbb{S}^{p}\). This implies that for these metric spaces, the polynomial uniform covering assumption in Theorem \ref{thm:ll-F} a) is satisfied and  the convergence rate of \(\hat{F}_{\omega, x}(t)\) is \((n/\log n)^{-2/5}\) uniform in \(x \in \mathcal{X}\), \(\omega \in \mathcal{M}\), and \(t \in \mathbb{R}^{+}\), which is optimal in the minimax sense and cannot be improved.

Next, we establish the convergence of the estimated distance profiles average transport costs \(\hat{\cat}(\omega\mid x)\) under \(\mathcal{N}(\epsilon,\mathcal{F}_1) \lesssim \epsilon^{-v}\). The convergence rate for  the case \(\log \mathcal{N}(\epsilon,\mathcal{F}_1) \lesssim \epsilon^{-v}\) is discussed in the Supplement.

\begin{theorem}\label{thm:CAT}
	Under Assumptions \ref{as:a1} - \ref{as:a5},  for the conditional profile average transport costs  estimator $\hat{C}(\omega\mid x)$ defined by \eqref{e:ll-l},
	\begin{itemize}
		\item[a).] If  $\mathcal{N}(\epsilon,\mathcal{F}_1)\lesssim  \epsilon^{-v}$ and $N(\epsilon,\mathcal{M},d)\lesssim\epsilon^{-v_1}$ for $v>0$ and $v_{1}>0$,
		$$\sup_{\omega\in\mathcal{M}}\sup_{x\in\tdomain}\left|\hat{\cat}(\omega\mid x)-\cat(\omega\mid x)  \right|=O\left(\sqrt{\frac{|\log h_n |+\log n}{n h_{n}}} +h_n^2 \right) \text{ a.s.}. $$
		\item[b).] If $ \mathcal{N}(\epsilon,\mathcal{F}_1)\lesssim  \epsilon^{-v}$ and $\log N(\epsilon,\mathcal{M},d)\lesssim\epsilon^{-v_1}$ for $v>0$ and $0<v_{1}<2$,
		$$\sup_{\omega\in\mathcal{M}}\sup_{x\in\tdomain}\left|\hat{\cat}(\omega\mid x)-\cat(\omega\mid x)  \right|=O\left(\sqrt{\frac{1}{n h_{n}^{1+v_1/2}}} +h_n^2 \right) \text{ a.s.}. $$
	\end{itemize}
\end{theorem}
By a similar argument as in Corollary \ref{c:F}, one can obtain the best convergence rate when  selecting the asymptotically optimal bandwidth sequence \(h_{n}\). 
Details on this are provided in Supplement. Unlike distance profiles, which are CDFs for which straightforward empirical estimates can be employed, the convergence rate of conditional profile average transport costs is influenced not only by the function class \(\mathcal{F}_{1}\) but also by the covering number of the metric space \(\mathcal{M}\). When \(\mathcal{M}\) is a compact subset of \(\mathbb{R}\) or the sphere \(\mathbb{S}^{p}\), the covering number \(N(\epsilon, \mathcal{M}, d)\) is less than or proportional to \(\epsilon^{-1}\). The inequality \(\log N(\epsilon, \mathcal{M}, d) \lesssim \epsilon^{-1}\) holds for most statistically relevant  metric spaces, such as  the space of phylogenetic trees \citep{lin2021total}. For the 2-Wasserstein space of distributions  on a compact subset of \(\mathbb{R}\) that are absolutely continuous with respect to the Lebesgue measure with smooth densities, the covering number also satisfies \(\log N(\epsilon, \mathcal{M}, d) \lesssim \epsilon^{-1}\) \citep{gao2009rate,dubey2022depth}.

The following  result demonstrates  the asymptotic conditional validity of the prediction sets constructed by Algorithm \ref{a:cf}. 

\begin{theorem}\label{thm:conf}
	Under Assumptions \ref{as:a6} , \ref{as:a1} - \ref{as:a5}, for the prediction set $\hat{C}_{\alpha}$ defined by Algorithm \ref{a:cf} using local linear estimates \eqref{e:ll-f}, \eqref{e:ll-l} and \eqref{e:ll-h},
	$$\prob\left( Y_{n+1}\in \hat{C}_{\alpha}(X_{n+1}) \mid X_{n+1} \right)\geq 1-\alpha+o_{P}(1) .$$
\end{theorem}

\section{Multivariate predictors}\label{s:mpd}

In the previous sections, we have established methodology and theory of the proposed conformal prediction method for the case of univariate predictors. For the case of multivariate predictors we consider \(\bm{X} \in \mathcal{X}\), where \(\mathcal{X}\) is a compact subspace of \(\mathbb{R}^d\) for a fixed \(d\) and note that the previously proposed  density- \citep{izbicki2022cd},  CDF-  \citep{chernozhukov2021distributional}  or  kernel-based methods \citep{lei2014distribution, lei2018distribution} to obtain a conformity score are subject to the curse of dimensionality.    To address this problem, we  employ  a single index Fr\'echet regression approach \citep{bhattacharjee2023single}. 
Throughout this section, we employ boldface for  multivariate vectors  to distinguish them from scalars.

Single index models are well established and  strike a balance between more restrictive linear models and fully nonparametric models that are hard to interpret and subject to the curse of dimensionality \citep{hall1989projection, ichimura1993semiparametric}. They  provide dimension reduction and thereby achieve convergence rates comparable to one-dimensional nonparametric regression, thus avoiding the curse of dimensionality. 
Various extensions of single index models have been proposed over the years \citep{zhou2008dimension, zhu2009distribution, chen2011single, ferraty2011estimation, jiang2011functional, kuchibhotla2020efficient} and more recently this approach has been extended to accommodate object responses \citep{bhattacharjee2023single}.   For an object response \(Y \in \mathcal{M}\) and a multivariate predictor \(\bm{X} \in \mathcal{X}\), a single index Fr\'echet regression model is given by 
\begin{equation}\label{e:sir}
	\E(Y\mid \bm{X}=\bm{x})=\E (Y\mid \bm{X}=\bm{x}^{\top}\bm{\theta}_{0} ):=m(t,\bm{\theta}_{0}),
\end{equation}
where \(\bm{\theta}_{0}\) is the true slope parameter, and \(m\) is the underlying regression function that depends on the multivariate predictors \(\bm{X}=\bm{x}\) only through the single index \(t=\bm{x}^{\top}\bm{\theta}_{0}\). 

We can then extend the definition of conditional distance  profiles, profile average transport costs, and profile scores in Section \ref{s:mth} to the multivariate case through
\begin{equation}\label{e:mdp}
	{F}_{m,\omega, x }(t) =\, \mathbb{P}(d(\omega,Y) \leq t \mid \bm{X}^{\top}\bm{\theta}_{0} =x ), \ \text{for all } t \in \mathbb{R}^{+},
\end{equation}
\begin{equation}\label{e:mpc}
	\cat_m(\omega\mid x)= \,\E \left[\int_{0}^{1} |F_{m,\omega,\bm{X}^{\top }\bm{\theta}_{0}}(t)- F_{m,Y, \bm{X}^{\top }\bm{\theta}_{0}}(t)|\,\diff t  \mid \bm{X}^{\top }\bm{\theta}_{0}=x\right] ,
\end{equation}
and 
\begin{equation}\label{e:mds}
	\cps_{m}(z\mid x):=\,\prob(\cat_{m}(Y\mid \bm{X}^{\top }\bm{\theta}_{0})\leq z\mid \bm{X}^{\top }\bm{\theta}_{0}=x ).
\end{equation}

Adopting the  estimation procedure of \cite{bhattacharjee2023single}, to obtain the slope vector \(\bm{\theta}_{0}\) one needs to estimate the conditional Fr\'echet mean \(m(t,\bm{\theta})\) for given \(\bm{\theta}\) by  
\begin{equation}\label{e:lf}
	\hat{m}(t,\bm{\theta}) =\,\, \argmin{\omega \in \mathcal{M}} \frac{1}{n} \sum_{i=1}^{n} \hat{s}(\bm{X}_{i}^{\top}\bm{\theta},t,h) d^{2}(Y_i,\omega),
\end{equation}
where 
\begin{equation}\label{e:fs}
	\hat{s}(\bm{X}_{i}^{\top}\bm{\theta},t,h) =\, \frac{1}{\hat{\sigma}_0^2(t,\bm{\theta})}\frac{1}{h} \K{\frac{\bm{X}_{i}^{\top}\bm{\theta}-t}{h}} \{\hat{\mu}_2(t,\bm{\theta})-\hat{\mu}_1(t,\bm{\theta})(\bm{X}_{i}^{\top}\bm{\theta}-t) \},
\end{equation}
with
\[
\hat{\mu}_{l}(t,\bm{\theta}) = \,\frac{1}{n} \sum_{j=1}^{n}\frac{1}{h} \K{\frac{\bm{X}_{j}^{\top}\bm{\theta}-t }{h}} (\bm{X}_{j}^{\top}\bm{\theta}-t )^l \quad \text{for } l=0,1,2,
\]
and \(\hat{\sigma}_0^2(t,\bm{\theta})=\hat{\mu}_2(t,\bm{\theta})\hat{\mu}_0(t,\bm{\theta})-\hat{\mu}_1^2(t,\bm{\theta})\). 
The parameter \(\bm{\theta}_{0}\) is then obtained  by minimizing the distance between \(Y_{i}\) and \(\hat{m}(\bm{X}_{i}^{\top}\bm{\theta},\bm{\theta})\). To ensure identifiability, \(\bm{\theta}\) is constrained to have  unit norm and to fall into the parameter space  
\[\Theta:= \{\bm{\theta} = (\theta_{1},\ldots,\theta_{d}) \in \mathbb{R}^d:\,\, \|\bm{\theta}\| = 1, \theta_{1} > 0\}.\]
The set \(\mathcal{X}_{\bm{\theta}_{0}}\) is defined as the image of \(\bm{x}^{\top}\bm{\theta}_{0}\), which is a compact subset of \(\mathbb{R}\) due to the compactness of \(\mathcal{X}\). 
Following  \cite{bhattacharjee2023single}, we partition \(\mathcal{X}_{\bm{\theta}}\) into \(M\) equal-width, non-overlapping bins \(\{B_{1},\ldots,B_{M}\}\) and  denote the representative data points in the \(l\)th bin as \((\tilde{\bm{X}}_{l},\tilde{Y}_{l})\), which satisfy \(\tilde{\bm{X}}_{l}^{\top}\bm{\theta} \in B_{l}\) for \(l = 1,\ldots,M\). The choice of the optimal $M$ depends on the metric space $\mathcal{M}$. For common  metric spaces, such as the 2-Wasserstein space, $M$ should be on the order of $n^{\gamma}$, where $0<\gamma<1/3$.  The final estimator for the true slope \(\bm{\theta}_{0}\) is then 
\begin{equation}\label{e:ht}
	\hat{\bm{\theta}}=\,\,\argmin{\bm{\theta}\in \Theta }\frac{1}{M}\sum_{l=1}^{M}d^{2}\left(\tilde{Y}_{l},\hat{m}(\bm{\tilde{X}}_{l}^{\top }\bm{\theta} ,\bm{\theta}) \right),
\end{equation}
where \(\hat{m}(\cdot, \cdot)\) is the estimator as defined in \eqref{e:lf}. 

We then implement the estimation procedure in Section \ref{s:ll-e} for the  data \((\bm{X}_{i}^\top \hat{\bm{\theta}}, Y_{i})\) and construct the prediction set \(\hat{C}_{\alpha}\) by Algorithm \ref{a:cf}. The local linear estimator for \(F_{m,\omega, x}\) is given by
\begin{equation}\label{e:ll-f-m}
	\hat{F}_{m,\omega, x}(t)=\,\,\argmin{\beta_{0}\in\reall}\frac{1}{nh_n}\sum_{j=1}^{n}\left\{ L_{j}(\omega,t)-\beta_{0}-\beta_{1}(\bm{X}_{j}^{\top }\hat{\bm{\theta}}-x) \right\}^2\K{\frac{\bm{X}_{j}^{\top }\hat{\bm{\theta}}-x}{h_n}},
\end{equation}
where \(L_{j}(\omega,t) = \mathds{1}_{\{d(\omega, Y_{j}) \leq t\}}\) as before. Subsequently, the conditional profile average transport costs are estimated by applying local linear smoothing for the  \(J_{m,j}(\omega,x) = \int_{0}^{1} |\hat{F}_{m,\omega, x}(t) - \hat{F}_{m,Y_{j}, \bm{X}_{j}^{\top}\hat{\bm{\theta}}}(t)| \, \diff t\) constructed with the estimated distance profiles \(\hat{F}_{m,\omega, x}\) as responses,   leading to
\begin{equation}\label{e:ll-l-m}
	\hat{\cat}_m(\omega\mid x)=\,\,\argmin{\beta_{0}\in\reall}\frac{1}{nh_n}\sum_{j=1}^{n}\left\{ J_{m,j}(\omega,x)-\beta_{0}-\beta_{1}(\bm{X}_{j}^{\top }\hat{\bm{\theta}}-x) \right\}^2\K{\frac{\bm{X}_{j}^{\top }\hat{\bm{\theta}}-x}{h_n}}.
\end{equation}
The estimate of the cumulative distribution function of \( C_{m}(\omega \mid x) \) emerges as
\begin{equation}\label{e:ll-h-m}
	\hat{\cps}_m(z\mid x)=\,\,\argmin{\beta_{0}\in\reall}\frac{1}{nh_n}\sum_{j=1}^{n}\left\{ H_{m,j}(z)-\beta_{0}-\beta_{1}(\bm{X}_{j}^{\top }\hat{\bm{\theta}}-x) \right\}^2\K{\frac{\bm{X}_{j}^{\top }\hat{\bm{\theta}}-x}{h_n}},
\end{equation}
where $H_{m,j}(z)=\mathds{1}_{\{\hat{\cat}(Y_{j}\mid \bm{X}^{\top}_{j}\hat{\bm{\theta}})\leq z \}}$.

To obtain asymptotic conditional validity,  continuity assumptions for \(F_{m,\omega, x}(t)\), \(C_{m}(\omega\mid x)\), and \(S_{m}(z\mid x)\) similar to those  in assumptions \ref{as:a4} - \ref{as:a5} are needed. Detailed  assumptions (B1)-(B5) are listed  in Supplement S4.

\begin{theorem}\label{t:mtc}
	Under Assumptions {(B1)} -- {(B5)} in Supplement S4,  if $h^{-1}\|\hat{\bm{\theta}}-\bm{\theta}_{0}\|=o_{P}(1)$, the prediction set $\hat{C}_{\alpha} $ obtained by Algorithm \ref{a:cf} with \eqref{e:ll-f-m} to \eqref{e:ll-h-m} satisfies 
		$$\prob(Y_{n+1}\in\hat{C}_{\alpha}\mid \bm{X}_{n+1}^{\top}\bm{\theta}_{0})\geq 1-\alpha+o_{P}(1). $$
\end{theorem}

Theorem \ref{t:mtc} demonstrates that the asymptotic conditional coverage is  guaranteed if \(\hat{\bm{\theta}}\) is a consistent estimator of \(\bm{\theta}_{0}\). Under certain regularity assumptions (Assumptions (U1) to (U8) in Supplement S4), Theorem 3.2 in \cite{bhattacharjee2023single}  implies that \(\|\hat{\bm{\theta}} - \bm{\theta}_{0}\| = O_{P}(M^{-1/2})\). Therefore the assumption $h^{-1}\|\hat{\bm{\theta}}-\bm{\theta}_{0}\|=o_{P}(1)$ in Theorem \ref{t:mtc} can be  satisfied by choosing the number of bins $M$ such that  \(M h^2\rightarrow\infty \).

\section{Simulations}\label{s:sim}
\subsection{Univariate predictors}\label{s:su}
We  illustrate the proposed method for univariate predictors with responses in various metric spaces, including the Euclidean space \(\mathbb{R}\), the sphere \(\mathbb{S}^2\), and the 2-Wasserstein space \(\mathcal{W}_2\). We use the conditional coverages and lengths (or sizes) of prediction sets as criteria. Unless otherwise specified,  for all settings the predictors \(x_i\) are generated from \(\text{Unif}(-1,1)\) and are independent of the regression error \(\epsilon_i\) in each setting.

\begin{figure}[t]
\centering
\includegraphics[width=\textwidth]{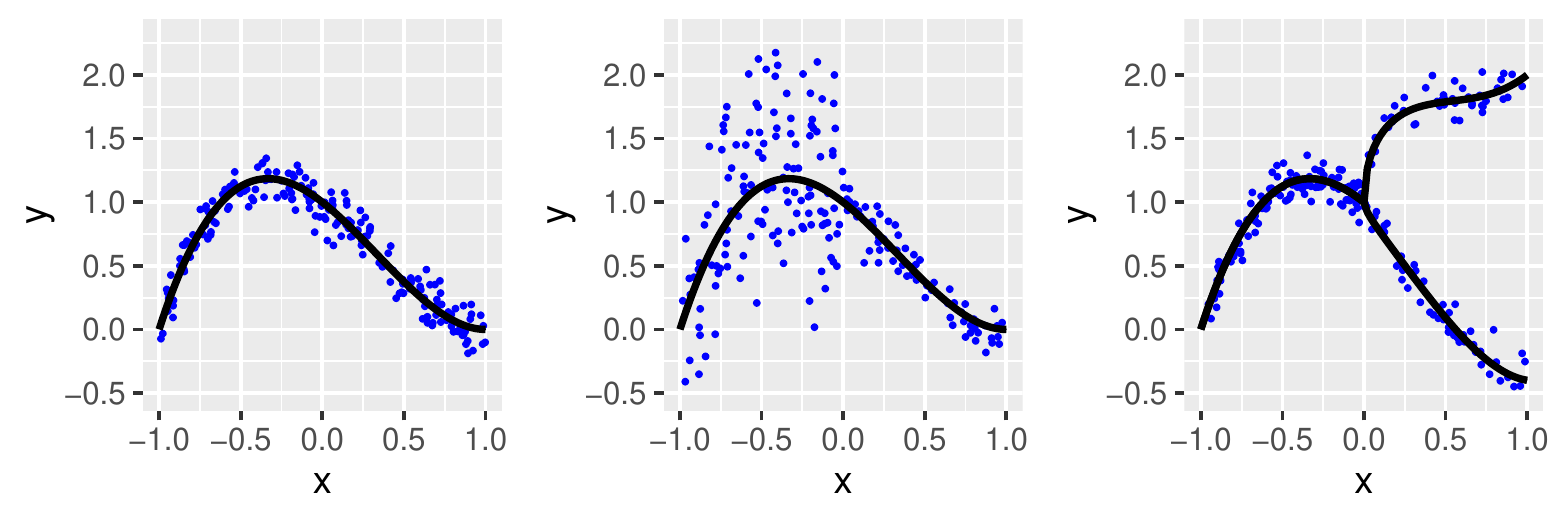}
\caption{Illustration of the  settings considered for $\mathcal{M}=\reall$: {Setting 1} (\emph{nonlinear regression with homoscedastic variability}, left panel); {Setting 2} (\emph{nonlinear regression with {heteroscedastic variability}}, middle panel); Setting 3 (\emph{nonlinear regression with bimodal pattern}, right panel).  Blue points are the observed data for the training set; the black curves are the underlying  regression functions or the mixture regression function (in the right panel).}
\label{fig:R1_settings}
\end{figure}

For  Euclidean responses, adopting  similar settings as in \cite{lei2014distribution} and \cite{izbicki2022cd}, we consider three scenarios  that include homoscedastic variability, heteroscedastic variability, and bimodal distributions of the responses, as  illustrated in Figure \ref{fig:R1_settings}.

\begin{itemize}
	\item \textbf{Setting 1} (\emph{Nonlinear regression with homoscedastic variability}):  This is a simple nonlinear regression scenario with homoscedastic errors.  The responses are generated by \(y_{i}=f(x_{i})+\epsilon_{i}\) with \(f(x)=(x-1)^2(x+1)\) and \(\epsilon_{i}\) are random samples from \(\mathcal{N}(0,0.1^2)\).
	\item \textbf{Setting 2} (\emph{Nonlinear regression with heteroscedastic variability}): The responses are generated by the same regression function as in {Setting 1}, but the regression errors have different variances for \(x_{i}\in(-1,0)\) and \(x_{i}\in(0,1)\), that is, \(y_{i}=f(x_{i})+\epsilon_{i}(x_{i})\) with \(f(x)=(x-1)^2(x+1)\) and \(\epsilon_{i}(x)\) are random samples from \(\mathds{1}_{\{-1\leq x\leq 0\}}\mathcal{N}(0,0.5^2)+\mathds{1}_{\{0<x\leq 1\}}\mathcal{N}(0,0.1^2)\).
	\item \textbf{Setting 3} (\emph{Nonlinear regression with a bimodal pattern}): We also consider a bimodal setting as considered previously in  \cite{lei2014distribution}. For \(x_{i}\in(-1,0)\), the regression function remains the same as in {Setting 1} and {Setting 2}. For \(x_{i}\in(0,1)\), two branches are {present}, each  with a probability of 0.5. Formally, the responses are generated by 
	\[y_{i} \sim 0.5 \mathcal{N}\left(f(x_{i})+g(x_{i}), 0.1^2\right)+0.5 \mathcal{N}\left(f(x_{i})-0.2g(x_{i}),0.1^2\right),\] where \(f(x)=(x-1)^2(x+1)\) and \(g(x)=2 \sqrt{x} \mathds{1}_{\{x \geq 0\}}\).
\end{itemize}

\begin{figure}[t]
\centering
\includegraphics[width=0.8\textwidth]{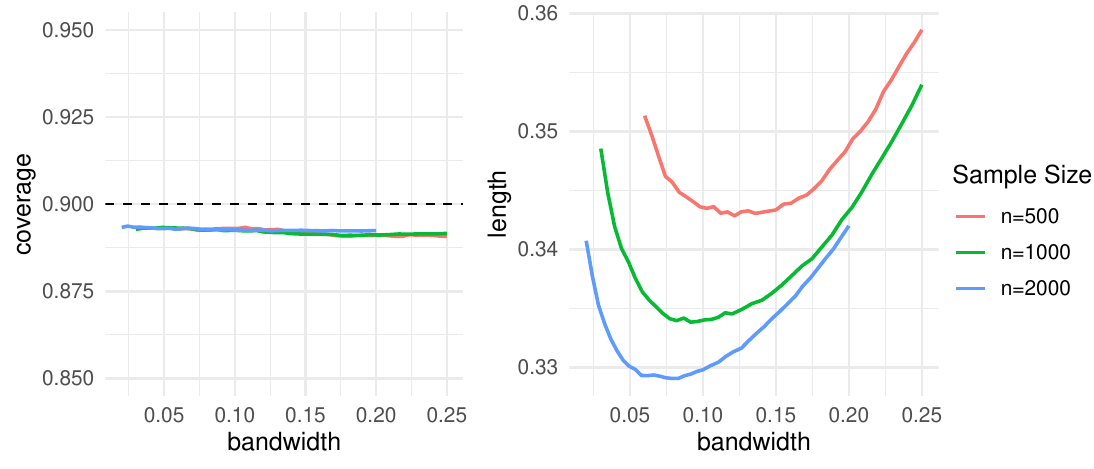}
\caption{Average  marginal coverage levels (left panel) and lengths of prediction sets (right panel) obtained with the proposed Conditional Profile Scores (CPS) over 200 Monte Carlo runs for varying bandwidths $h$ for {Setting 1} (\emph{nonlinear regression with homoscedastic variability}). The target coverage level is $90\%$. }
\label{f:m-1}
\end{figure}

We first check the influence of bandwidth choice on marginal coverage level and average length of the prediction sets. We considered sample sizes $n= 500, 1000, 2000$ and 200 Monte Carlo runs for each setting. Conditional coverage was evaluated on a test set with a sample size of 2000 with  the same distribution as the training set for each setting. Marginal coverage levels and lengths of prediction sets for {Setting 1} (\emph{Nonlinear regression with homoscedastic variability}) are shown in Figure \ref{f:m-1} and for {Setting 2} (\emph{Nonlinear regression with heteroscedastic variability}) and {Setting 3} (\emph{Nonlinear regression with a bimodal pattern})  in  Supplement S6. One key feature of conformal inference is that the choice of conformity scores does not affect the coverage level but  does affect the size (length) of the  prediction sets. This is verified in Figure \ref{f:m-1}. Due to the bias and variance trade-off for the local linear smoother, the length of the prediction set as a function of the bandwidth is convex.  As the sample size increases, the lengths of the conformal sets and the optimal bandwidths decrease, consistent with theory.

Next we compare conditional coverage levels and lengths of the conformal prediction sets for the proposed  conditional profile scores (CPS) and previously established conformity scores,  including the HPD-split scores proposed in \cite{izbicki2022cd} and  Conformalized Quantile Regression (CQR) in \citep{romano2019conformalized, sesia2020comparison}. The HPD-split method was implemented using the R code available at \url{https://github.com/rizbicki/predictionBands}, and the CQR methods using the Python code available at \url{https://github.com/msesia/cqr-comparison}; these methods were implemented with  their default settings.

\begin{figure}[t]
\centering
\includegraphics[width=.99\textwidth]{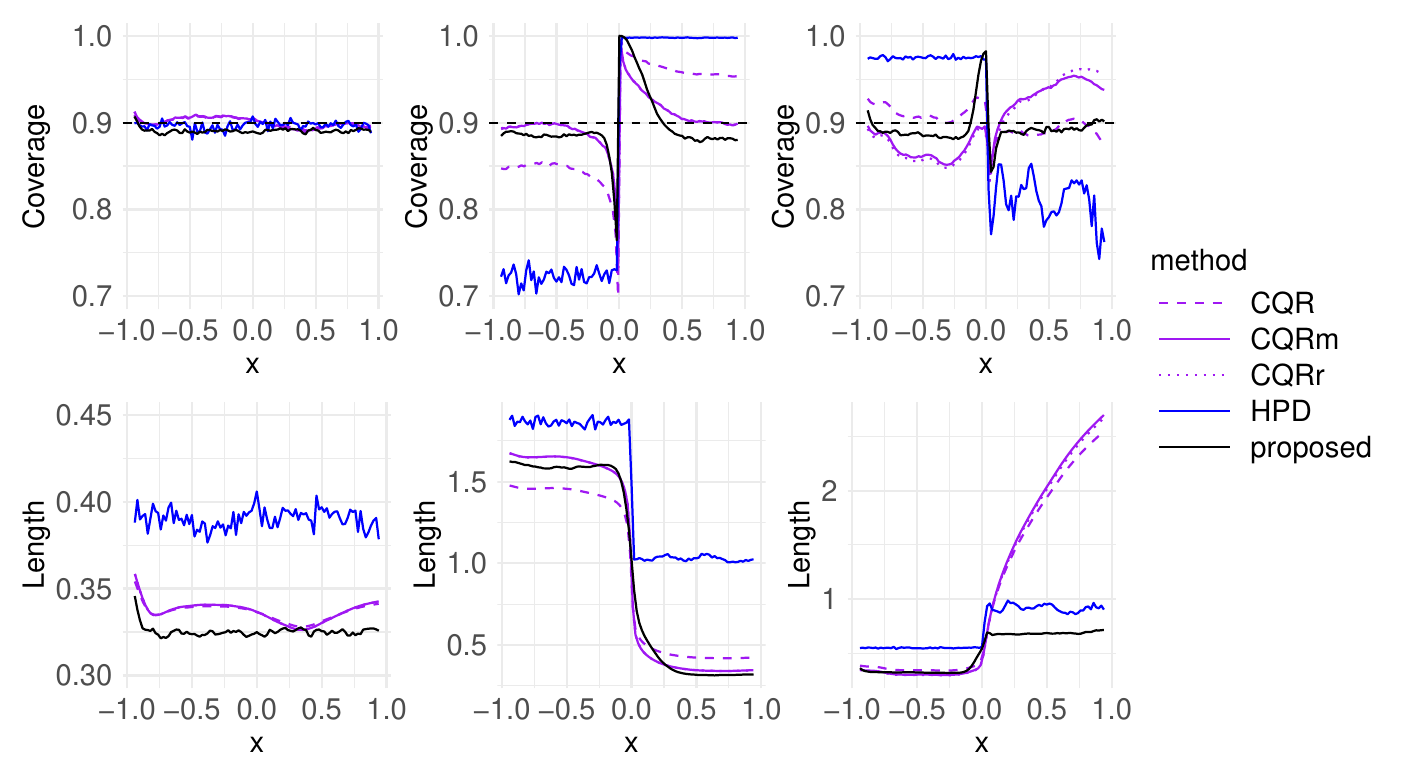}
\caption{Average conditional coverage (first row) and prediction set length (second row) over 200 Monte Carlo runs in dependence  on the level of the predictor \(x\). The same three settings for \(\mathcal{M}=\reall\) as in Figure \ref{fig:R1_settings} are considered, with a sample size of $n=2000$ and a target coverage level of 90\%. The left column corresponds to Setting 1 (\emph{nonlinear regression with homoscedastic variability}), the middle column to Setting 2 (\emph{nonlinear regression with heteroscedastic variability}), and the right column to Setting 3 (\emph{nonlinear regression with a bimodal pattern}). The prediction sets are obtained using Algorithm \ref{a:cf}. Results for the conditional profile scores (proposed) are shown in black,  for the HPD-split method  \citep{izbicki2022cd} in blue and  for several variants of the CQR method \citep{romano2019conformalized, sesia2020comparison} in purple, with versions CQRm (solid purple),   CQR (dashed purple)  and CQRr (dotted purple).  All methods are run using the default setting as provided in the respective code.}
\label{f:c-1}
\end{figure}

As illustrated in the first row of Figure \ref{f:c-1}, the proposed method consistently achieves conditional coverage across all settings. While the HPD-split method is theoretically expected to achieve conditional validity, in {Setting 2} (\emph{nonlinear regression with heteroscedastic variability}) and {Setting 3} (\emph{nonlinear regression with a bimodal pattern}), where there is a change point in   variance and mean at \(x=0\), the HPD-split shows varying coverage levels for \(x \in (-1, 0)\) and \(x \in (0, 1)\) and only achieves  marginal coverage. This is due to the inaccurate estimation of conditional density functions in this complex setting; further details can be found in the Supplement.  In contrast, conditional profile scores generally achieve conditional validity for all three settings. The spike at \(x=0\) is caused by the change point. 
The second row of Figure \ref{f:c-1} reveals that the proposed method results in prediction sets with shorter lengths compared to the HPD-split in all three settings. Compared to the CQR methods, the proposed scores have similar lengths in {Setting 1} and {Setting 2}, but much smaller lengths in  {Setting 3}. These results demonstrate the efficiency of conditional profile scores. Additional simulation results and the comparison with Distributional Conformal Prediction \citep{chernozhukov2021distributional} can be found in the Supplement S6.1.

Next we consider responses in metric spaces, specifically responses on the unit sphere \(\mathbb{S}^2:=\{p \in \mathbb{R}^{3} \mid  p^{\top}p = 1\}\) and in the 2-Wasserstein space. Note that \(\mathbb{S}^2\) is a 2-dimensional Riemannian manifold endowed with the geodesic distance \(d(p,q) = \arccos(p^{\top}q)\). The tangent space at a point \(p\) is \(T_{p}:=\{y \in \mathbb{R}^{3} \mid  y^{\top}p = 0\}\). For all \(p \in \mathbb{S}^2\) and \(v \in T_{p}\), the Riemannian exponential map that projects \(v\) onto \(\mathbb{S}^2\) is defined by 
\[\exp_{p}v = \cos(\|v\|)p + \sin(\|v\|)\|v\|^{-1}v.\]

\begin{itemize}
	\item \textbf{Setting 4} (\emph{Responses in the unit sphere  \(\mathbb{S}^2\)}). The responses are generated by \[y_{i} = \exp_{\mu(x_{i})}V_{i}(x_i),\] with \(\mu(x) = (\sin(\pi x/2), \cos(\pi x/2), 0)^{\top}\) and \(V_i = (0, 0, \epsilon_i)^{\top}\), where \(\epsilon_{i} \sim_{i.i.d.}  \mathcal{N}(0, 0.5^2)\).
\end{itemize}
For the next setting with distributional data in the  2-Wasserstein space, we adopt  addition and scalar multiplication operations in the transport space \(\mathfrak{T}:=\{T:[0,1]\mapsto[0,1],\ T(0)=0,\ T(1)=1, \ T \text{ is increasing}\}\)  following  \cite{zhu2021autoregressive},  as follows, 
\begin{itemize}
	\item Addition: \(T_{1}\oplus T_{2}=T_{2}\circ T_1\) for \(T_1, T_2 \in \mathfrak{T}\).
	\item Scalar multiplication: for any \(|\alpha|\leq 1\) and \(T \in \mathfrak{T}\),
	\[
	\alpha \odot T(x):=\,\begin{cases}
	x+\alpha(T(x)-x), & 0<\alpha \leq 1, \\
	x, & \alpha=0, \\
	x+\alpha\left(x-T^{-1}(x)\right), & -1 \leq \alpha<0.
	\end{cases}
	\]
\end{itemize}
For distributions defined on \((0,1)\) that are absolutely continuous with respect to the Lebesgue measure their corresponding quantile functions can be regarded as elements of  \(\mathfrak{T}\). For the 2-Wasserstein space, we represent the  random elements in \(\mathcal{W}_2\) through their quantile functions. 

\begin{figure}[t]
\centering
\includegraphics[width=.85\textwidth]{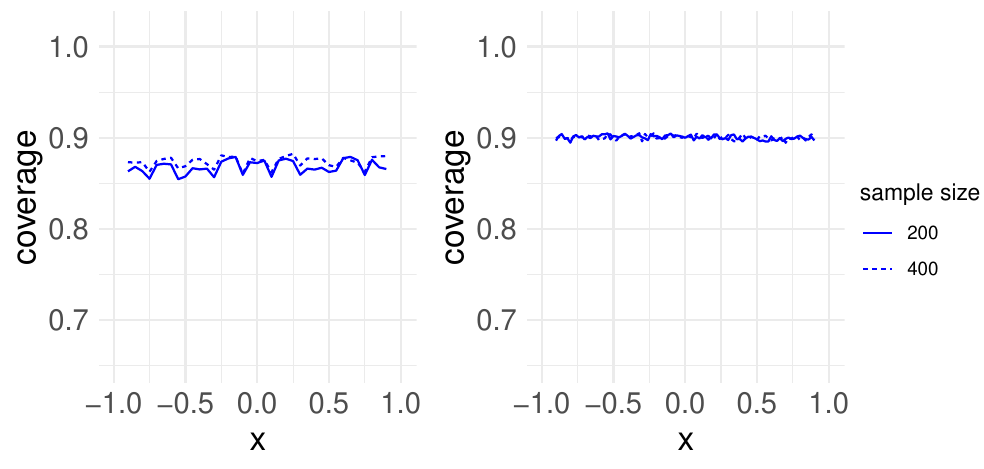}
\caption{Average  conditional coverage levels over 200 Monte Carlo runs for the proposed CPS conformal method and a  target coverage level at $90\%$ in dependence on the level of the predictor, for Setting 4 (\emph{unit sphere  \(\mathbb{S}^2\)}, left panel) and Setting 5 (\emph{Wasserstein space \(\mathcal{W}_2\)}, right panel).} 
\label{f:c_SW}
\end{figure}

\begin{itemize}
	\item \textbf{Setting 5} (\emph{Distributional responses in the Wasserstein space \(\mathcal{W}_2\)}).  The responses are $y_{i}=\operatorname{Trun}\mathcal{N}(f(x_{i}),0.5)\oplus \epsilon_{i}$, where \(\operatorname{Trun}\mathcal{N}\) is the truncated normal distribution on \((0,1)\), \(f(x_{i})=0.8(x_{i}-1)^2(x_{i}+1)\), and the  \(\epsilon_{i}\) are random distributions drawn from  \newline  \(\operatorname{Unif}[-0.5,0.5]\odot \mathrm{Beta}(2,2)\).
\end{itemize}
The conditional validity of   the proposed conditional profile scores for {Setting 4} and {Setting 5} is demonstrated 
in Figure \ref{f:c_SW}. Additional simulation results, such as marginal coverage levels and  sizes of the prediction sets can be found  in Supplement S6.

\subsection{Multivariate predictors}\label{s:sm}

In this subsection, we show the performance of the proposed method described in Section \ref{s:mpd} for multivariate predictors and scalar responses.  In addition to evaluating the conditional coverage and size of the prediction set, we also examine the mean square error of the slope parameter in the single index Fr\'echet regression, 
\begin{equation}\label{e:mse}
	\operatorname{MSE}(\bm{\theta_{0}},\hat{\bm{\theta}}) = \|\hat{\bm{\theta}} - \bm{\theta_{0}}\|^2.
\end{equation}
For responses we considered the same scenarios as for univariate responses  and again  compared the proposed   conditional profile scores with  HPD-split scores \citep{izbicki2022cd} for  Euclidean responses across three different settings, as well as for responses located on the unit sphere. 
\begin{itemize}
	\item \textbf{Setting 6} (\emph{Multivariate predictor with homoscedastic variability}): The predictors are \(\bm{X}_{i}=(x_{i1},x_{i2})^{\top}\) with \(x_{ik}\) \(i.i.d.\) \(\sim \mathrm{Unif}(-1,1)\) and \(\bm{\theta_{0}}=(1,0)^{\top}\). The responses are generated by \(y_{i}=f(\bm{X}_{i}^{\top}\bm{\theta_{0}})+\epsilon_{i}\) with \(f(x)=(x-1)^2(x+1)\) and the \(\epsilon_{i}\) are random samples from \(\mathcal{N}(0,0.1^2)\).
	\item \textbf{Setting 7} (\emph{Multivariate predictor with heteroscedastic variability}): The predictors are \(\bm{X}_{i}=(x_{i1},x_{i2})^{\top}\) with \(x_{ik}\) \(i.i.d.\) \(\sim \mathrm{Unif}(-1,1)\) and \(\bm{\theta_{0}}=(1,0)^{\top}\). The responses are generated by \(y_{i}=f(\bm{X}_{i}^{\top}\bm{\theta_{0}})+\epsilon_{i}(\bm{X}_{i}^{\top}\bm{\theta_{0}})\) with \(f(x)=(x-1)^2(x+1)\) and the  \(\epsilon_{i}(x)\) are random samples from \(\mathds{1}_{-1\leq x\leq 0}\mathcal{N}(0,0.5^2)+\mathds{1}_{0<x\leq 1}\mathcal{N}(0,0.1^2)\).
	\item \textbf{Setting 8} (\emph{Multivariate predictor with a bimodal pattern}): The predictors are \(\bm{X}_{i}=(x_{i1},x_{i2})^{\top}\) with \(x_{ik}\) \(i.i.d.\) \(\sim \mathrm{Unif}(-1,1)\) and \(\bm{\theta_{0}}=(1,0)^{\top}\), and  responses are  
	\[y_{i} \sim 0.5 \mathcal{N}\left(f(\bm{X}_{i}^{\top}\bm{\theta_{0}})+g(\bm{X}_{i}^{\top}\bm{\theta_{0}}), 0.1^2\right)+0.5 \mathcal{N}\left(f(\bm{X}_{i}^{\top}\bm{\theta_{0}})-0.2g(\bm{X}_{i}^{\top}\bm{\theta_{0}}),0.1^2\right),\] where \(f(x)=(x-1)^2(x+1)\) and \(g(x)=2\sqrt{x} \mathds{1}_{\{x \geq 0\}}\).
\end{itemize}

\begin{figure}[t]
\centering
\includegraphics[width=.86\textwidth]{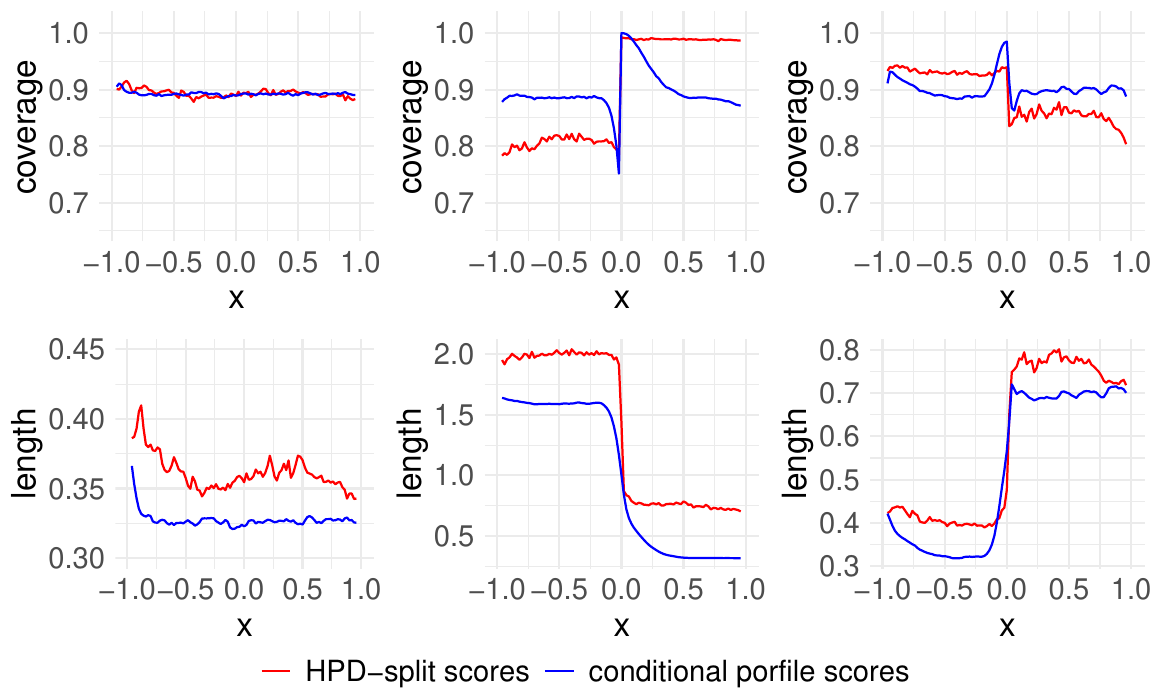}
\caption{Average conditional coverages (first row) and prediction set lengths (second row) over 200 Monte Carlo runs in dependence on the single index level $x$ for multivariate predictors in Settings 6 (\emph{Multivariate predictor with homoscedastic variability}, left column),  7 (\emph{Multivariate predictor with heteroscedastic variability}, middle column) and  8 (\emph{Multivariate predictor with a bimodal pattern}, right column) 
for  sample size  \(n=2000\) and target coverage level  90\%. The prediction sets are obtained by  Algorithm \ref{a:cf} as in  \eqref{e:ll-f-m} -- \eqref{e:ll-h-m}; results in blue are for the  proposed  conditional profile scores and those in red for the HPD-split scores  \citep{izbicki2022cd}. }
\label{f:c-d2-1}
\end{figure}

Figure \ref{f:c-d2-1} demonstrates conditional coverages and lengths of prediction sets, in analogy to  Figure \ref{f:c-1}. Conditional profiles scores  outperform HPD-split scores in both coverage and size. For  responses on the unit sphere  with a multivariate predictor, we consider the following setting. Figure S.10 in Supplement S6.2 demonstrates  that the proposed Fr\'echet single index approach with Algorithm \ref{a:cf} achieves conditional validity for this setting ({Setting 9}).
\begin{itemize}
	\item \textbf{Setting 9} (\emph{Multivariate predictor with responses in \(\mathbb{S}^2\)}): The predictors are \[\bm{X}_{i}=(x_{i1}, x_{i2}, x_{i3}, x_{i4})^{\top}\] with \(x_{ik}\) independently and identically distributed  \(\sim \mathrm{Unif}(-1,1)\) and \(\bm{\theta_{0}}=(1,0,0,0)^{\top}\). The responses are generated by \(y_{i}=\exp_{\mu(\bm{x_{i}}^{\top}\bm{\theta_{0}})}V_{i}(x_i)\), where \[\mu(x)=(\sin(\pi x/2), \cos(\pi x/2), 0)^{\top}\] and \(V_i=(0, 0, \epsilon_i)^{\top}\) where the  \(\epsilon_{i}\) are  random samples from \(\mathcal{N}(0,0.5^2)\).
\end{itemize}

We also obtained   \(\operatorname{MSE}(\hat{\bm{\theta}}, \bm{\theta_0})\) as defined in \eqref{e:mse} for {Settings 6-9}  across various sample sizes, as this error affects the estimation of the proposed conditional profile score when one uses single index Fr\'echet regression.  The results in Table \ref{t:mse} indicate that  \(\operatorname{MSE}(\hat{\bm{\theta}}, \bm{\theta_0})\)  decreases as the sample size increases across all settings so that this error will be small when one has large enough sample sizes.

 \begin{table}[t]
  \centering
  \caption{Average \(\operatorname{MSE}(\hat{\bm{\theta}},\bm{\theta_0})\) for the estimated single index parameter $\hat{\bm{\theta}}$ over 200 Monte Carlo runs for various settings, with standard deviations in parentheses, where all values are multiplied by \(10^2\) for better visualization  }
    \begin{tabular}{rlllrrrrr}\hline
\cmidrule{1-4}\cmidrule{6-9}          & \multicolumn{3}{c}{$\mathcal{M}=\reall$} &       & \multicolumn{4}{c}{$\mathcal{M}=\mathbb{S}^2$} \\
\cmidrule{1-4}\cmidrule{6-9}    \multicolumn{1}{l}{$n$} & \textbf{Setting 6} & \textbf{Setting 7} & \textbf{Setting 8} &       & \multicolumn{1}{l}{$n$} & \multicolumn{1}{l}{\textbf{Setting 9}} &       &  \\
\cmidrule{1-4}\cmidrule{6-9}    500   & 0.52(0.45) & 2.85(2.71) & 20.23(32.50) &       & 200   & \multicolumn{1}{l}{12.84(43.21)} &       &  \\
\cmidrule{1-4}\cmidrule{6-9}    1000  & 0.38(0.30) & 1.82(1.86) & 8.05(21.35) &       & 500   & \multicolumn{1}{l}{7.58(34.01)} &       &  \\
\cmidrule{1-4}\cmidrule{6-9}    2000  & 0.24(0.23) & 1.22(1.14) & 2.42(2.48) &       &       &       &       &  \\
\cmidrule{1-4}\hline    \end{tabular}%
  \label{t:mse}%
\end{table}%

\section{Data illustrations}\label{s:rda}
\subsection{New York taxi data}

Trip records for yellow taxis in New York City, with times and locations for  pick-ups and drop-offs,  can be accessed via \url{https://www.nyc.gov/site/tlc/about/tlc-trip-record-data.page}. We focus on the pick-up and drop-off points located within Manhattan. Omitting Governor's Island, Ellis Island and Liberty Island, we divide the remaining 66 zones of Manhattan into 13 distinct regions. The predictor \(x\) records the time of day, ranging from 4 AM to 8 PM and the   response is a network representing the number of customers commuting between the selected  areas by taking a yellow taxi at time \(x\); we include  all $N=260$ weekdays within the year 2023. For the $i$th weekday, there are  \(n_i\) taxi trips that take place  between 4 AM and 8 PM. We divide the time domain \((4, 20)\) (corresponding to the time interval  4 AM to 8 PM) into bins \(S_{i1}=(a_{i0}, a_{i1}), S_{i2}=(a_{i1}, a_{i2}), \ldots, S_{B_i}=(a_{i(B_i-1)}, a_{iB_i})\), ensuring there are \(M=1000\) records within each bin, with \(4=a_{i0}<a_{i1}<a_{i2}<\cdots< a_{i(B_i-1)}<a_{iB_i}=20\) and \(B_i=\lfloor n_i/M \rfloor\). For each bin \(S_k\), \(k=1, \ldots, B_i\), we pool all records whose pickup times fall within this bin to construct the 13 by 13 adjacency matrix \(y_{ik}\), which represents the response at time \(x_{ik}=(a_{i(k-1)}+a_{ik})/2\).  Each adjacency matrix's edge weights are normalized against its maximum edge weight, ensuring they range between \([0, 1]\). The resulting  pairs \(\{(x_{ik}, y_{ik})\}_{k=1}^{B_i}\) from all weekdays are pooled together to form the final dataset.

\begin{figure}[tb]
\centering
\includegraphics[width=0.35\textwidth]{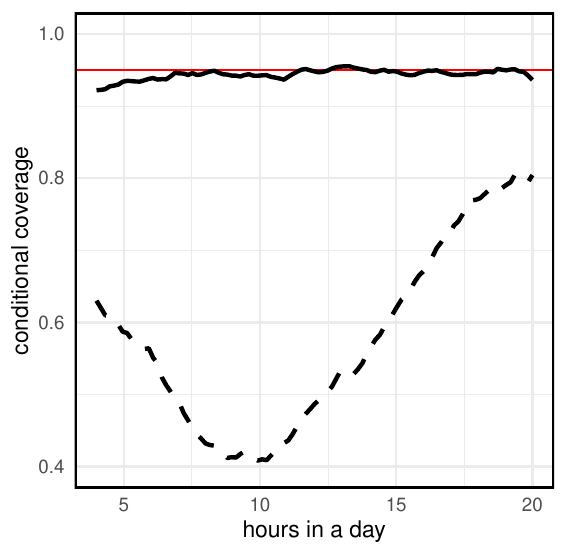}
\caption{Conditional coverage levels for taxi data. The target coverage level is 95\%, as indicated by the red solid line. The conditional coverage levels evaluated on the testing set (black solid line) derived from weekdays differs substantially from the   conditional coverage levels obtained when using data from weekends and holidays (dashed line).}
\label{f:taxi-cd}
\end{figure}

We use the Frobenius metric \(d_{F}\) as the metric between graph adjacency matrices, 
\[d_F(\mathbf{A}, \mathbf{B}) = \sqrt{\operatorname{tr}\left[(\mathbf{A} - \mathbf{B})(\mathbf{A} - \mathbf{B})^\top\right]},\]
for \(\mathbf{A}, \mathbf{B} \in \mathbb{R}^{13 \times 13}\). The data are  divided into training, calibration, and testing sets in a 4:4:2 proportion. We implemented Algorithm \ref{a:cf}  and evaluated  the conditional coverage on the testing set. Figure \ref{f:taxi-cd} indicates  that the proposed conditional profile score ensures  conditional coverage across all \(x\) in the time range. 
We also examined the conditional coverage for holidays and weekends in 2023, but still using training and calibration data collected for weekdays in Algorithm \ref{a:cf}. Figure \ref{f:taxi-cd} reveals that the conditional coverages for holidays significantly deviate from the target, confirming  that taxi transportation patterns on weekdays and non-weekdays do not align.

\begin{figure}[tb]
\centering
\includegraphics[width=\textwidth]{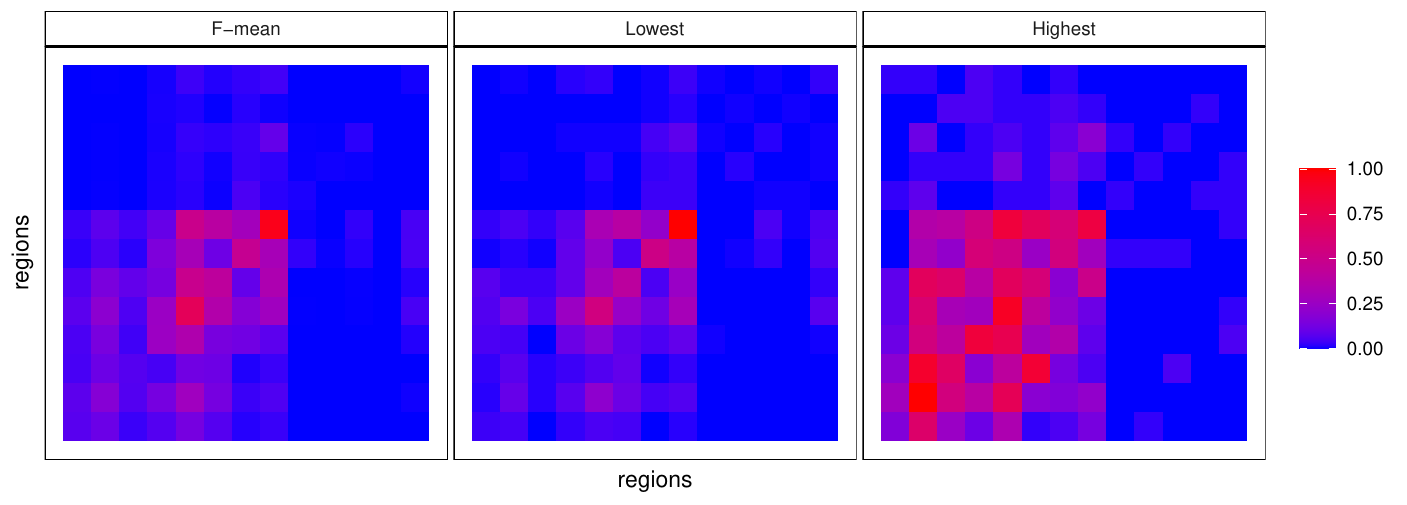}
\caption{Heat maps for networks represented by graph adjacency matrices at time 3 PM  from the training set of weekday data. The Fr\'echet mean is in the left panel,  the network  with the lowest conditional profile score in the middle panel and  the network  with the highest conditional profile score in the right panel.  }
\label{f:taxi-hm}
\end{figure}

Figure \ref{f:taxi-hm} displays  heatmaps for the Fr\'echet mean and for networks with the lowest and highest conditional profile scores from the training set. The heatmap for the network with the lowest score has a pattern similar to the Fr\'echet mean, indicating its corresponding adjacency matrix is at the center of the dataset. The heatmap for the network with the highest score presents a very  different pattern from the previous two, indicating that it sits near the boundary of the prediction set and has a higher likelihood of being an outlier.

\subsection{U.S. energy data}

The global energy landscape has undergone profound changes over the last thirty years, driven by technological innovations, economic shifts, and evolving societal needs.  Data on the sources of energy used for electricity generation across the U.S. are available at \url{https://www.eia.gov/electricity/data/state/}. As an illustration of the proposed method, we considered three categories of energy sources and their corresponding proportions: \RN{1}. Coal, Petroleum, Wood, and Wood Derived Fuels; \RN{2}. Natural Gas; \RN{3}. Hydroelectric, Wind, Nuclear, Geothermal, Solar Thermal and Photovoltaic. Sources in category \RN{1} are traditional energy sources known to emit high levels of greenhouse gases and have historically been associated with air pollution. Sources in category \RN{2} are  cleaner alternatives and their contribution has steadily grown.  Sources in category \RN{3} represent renewable energy and other  eco-friendly sources.

The predictors \(x\) are calendar years ranging from 1990 to 2021. The corresponding responses are defined as \(y(x) = (U^{1/2}(x), V^{1/2}(x), W^{1/2}(x))\), where \(U(x)\), \(V(x)\), and \(W(x)\) denote the proportions of energy sources \RN{1}, \RN{2}, and \RN{3} used for electricity generation in the given year \(x\). The proportions constitute compositional data, as they are non-negative and constrained by \( U(x) + V(x) + W(x) = 1 \) for each calendar year \( x \). Consequently, their square roots lie on the sphere \( \mathbb{S}^2 \). We then use the geodesic on the unit sphere as the metric.

\begin{figure}[t]
    \centering
    \begin{subfigure}{0.27\textwidth}
        \includegraphics[width=\linewidth]{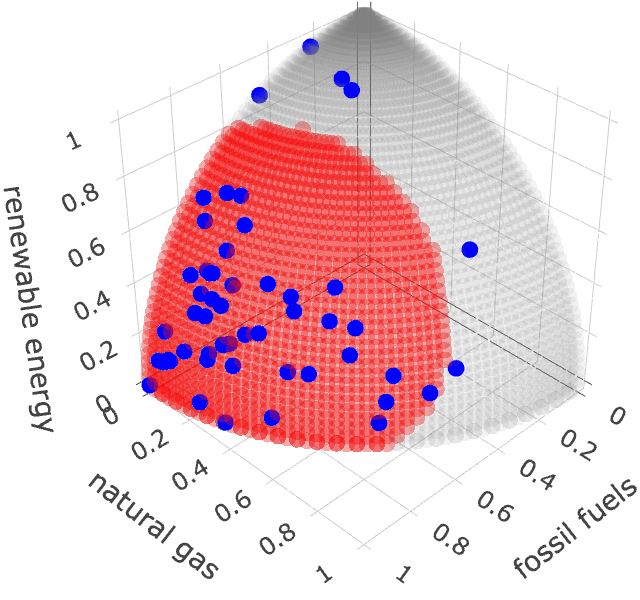}
        \caption*{1990}
    \end{subfigure}
    \begin{subfigure}{0.27\textwidth}
        \includegraphics[width=\linewidth]{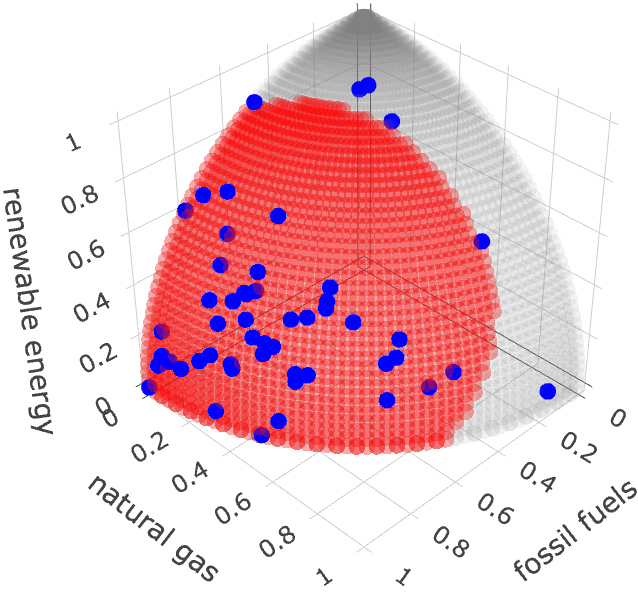}
        \caption*{1996}
    \end{subfigure}
    \begin{subfigure}{0.27\textwidth}
        \includegraphics[width=\linewidth]{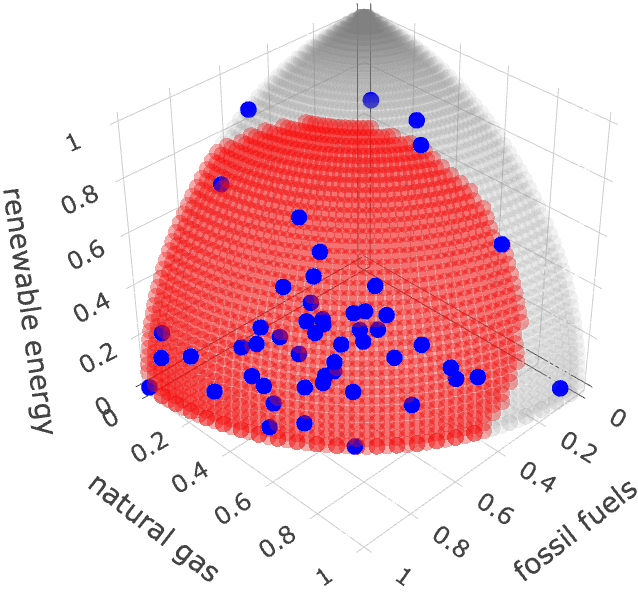}
        \caption*{2002}
    \end{subfigure}
    \begin{subfigure}{0.27\textwidth}
        \includegraphics[width=\linewidth]{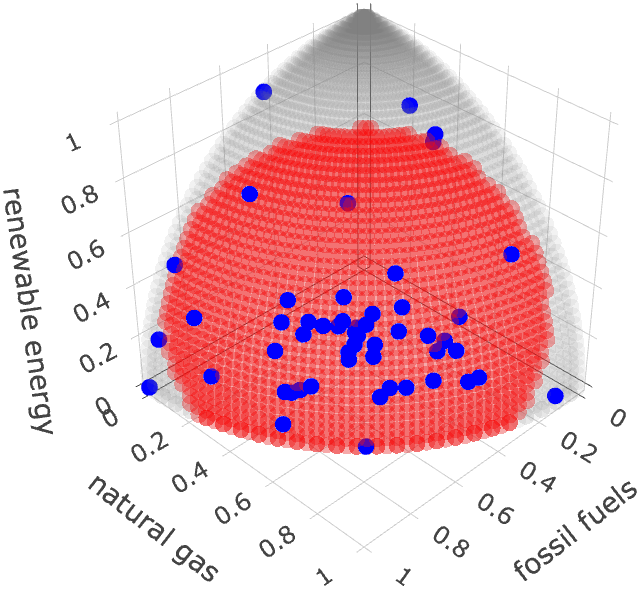}
        \caption*{2008}
    \end{subfigure}
    \begin{subfigure}{0.27\textwidth}
        \includegraphics[width=\linewidth]{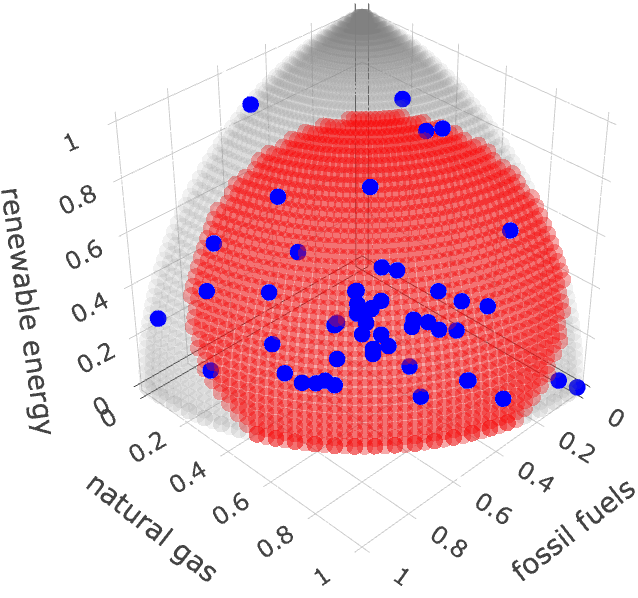}
        \caption*{2014}
    \end{subfigure}
    \begin{subfigure}{0.27\textwidth}
        \includegraphics[width=\linewidth]{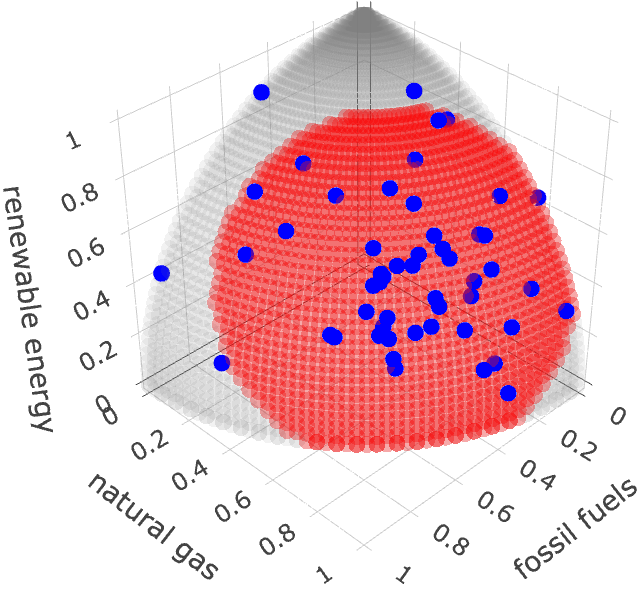}
        \caption*{2021}
    \end{subfigure}
    \caption{Illustration of energy data sources represented as data on the sphere $\mathbb{S}^2$ and the corresponding conformal sets for calendar years 1990, 1996, 2002, 2008, 2014, and 2021, where calendar year is the predictor.  The  front-right axis represents the proportion of fossil fuels (sources \RN{1}), the front-left axis represents the proportion of natural gas (sources \RN{2}) and the rear axis represents the proportion of renewable energy (sources \RN{3}).  Blue points represent the vector of square roots of  the proportions of three energy sources for each state. The red areas are the conformal prediction sets obtained with the proposed conditional profile score.}
    \label{f:eng}
\end{figure}

Figure \ref{f:eng} shows a clear trend in the prediction set obtained by the proposed method, moving from the bottom left to the top right as the years progress. This indicates a decreasing dependence on traditional fossil fuels and an increasing share of natural gas and renewable energy sources.

\section{Discussion}\label{s:dis}

We  extend the concept of distance profiles \citep{dubey2022depth} to a conditional version and introduce  the novel notion of profile average transport costs to quantify the conformity of any element in a metric space with respect to the underlying conditional distribution of \(Y\mid X\). 
While transport ranks account for the directionality of optimal transports by accounting for mass being transported to the left or to the right,   profile average transport costs focus solely on the costs of transports between two distance profiles and ignore  the direction of the transport.  Consequently, while transport ranks identify the most centrally located element globally, the proposed CPCs are not directly connected to centrality but can capture local modes with respect to the underlying conditional distributions, aiding in the construction of accurate  conformal prediction sets.

The  key for successful conformal inference lies in the choice of a good conformity score.  In general metric spaces, residual scores \(\hat{R}(x,y) = d(\hat{f}(x),y)\) may seem to be the most straightforward approach, however for complex object data these have many shortcomings. First, such  residual scores can only achieve marginal coverage, and the size of the resulting prediction sets is the same for all predictor levels.  This results in poor prediction sets when there is heteroscedasticity or another kind of distributional change in  \(Y\) when predictors vary.   Moreover,  residual scores depend crucially on the regression function estimator \(\hat{f}\). In many situations, including when the conditional distribution of \(Y\mid X\) is bimodal or multimodal, estimates \(\hat{f}\) will not perform well. 

 We reiterate that the  proposed conditional profile transport cost is not intended as a  centrality measure and a  most central point, for example an element of $\mathcal{M}$ with maximal transport rank, may not have the lowest CPC value. Instead, as we show in this paper,  it serves as the basis for a good  conformity measure for constructing prediction sets. Specifically, the CPS, i.e., the  CPC-based conformity scores, perform effectively across a wide range of settings and models, whereas centrality measure-based scores are only efficient for unimodal distributions. In cases where the underlying distribution is not centered around a single element, such as in the bimodal case,  prediction sets obtained using centrality measures such as  transport ranks  result in a single set centered at the global center of the data and thus are large and less informative. In such cases, the proposed CPC-based conformity scores produce smaller and more informative  prediction sets.  In case the conditional distribution is unimodal in a suitably defined sense,  the most central point can be included in the prediction sets based on CPC-based scores for reasonably large coverage levels. 

We further note that the  proposed conditional profile scores are determined solely by the underlying conditional probability measure and distances between random objects generated by this measure in  the metric space, leading to  an intrinsic approach that does not require projections of the  data to an extrinsic space, e.g., a tangent bundle.  This distance-based approach simplifies computations, as it can be readily applied for any metric space without the need to devise suitable transformations or manipulations of the data structure. The effectiveness of the proposed method is demonstrated through comparison with other conformal methods \citep{romano2019conformalized, sesia2020comparison, chernozhukov2021distributional, izbicki2022cd} for the case of Euclidean responses. In this special case the proposed method is found to perform equally well or  better in terms of both  conditional coverage accuracy and prediction set lengths.

 In future research, it will be of interest to determine how the proposed CPS compares with yet to be developed alternative conformity scores when responses are random objects in general metric spaces. Especially the  exploration of conformal prediction regions to address the construction of normal ranges or tolerance regions for the case of multivariate responses likely will have many applications, especially in medicine and  life sciences.  

\section*{Funding}
This research is supported in part by  NSF Grant DMS-2310450.

%
%
	
	\bigskip
	\bibliographystyle{Chicago}
	\bibliography{report.bib}

\begin{thebibliography}{}

\bibitem[\protect\citeauthoryear{Ambrosio, Gigli, and Savar{\'e}}{Ambrosio
  et~al.}{2008}]{ambrosio2008}
Ambrosio, L., N.~Gigli, and G.~Savar{\'e} (2008).
\newblock {\em Gradient Flows: in Metric Spaces and in the Space of Probability
  Measures}.
\newblock Springer Science \& Business Media.

\bibitem[\protect\citeauthoryear{Angelopoulos, Cand{\`e}s, and
  Tibshirani}{Angelopoulos et~al.}{2024}]{angelopoulos2023conformal}
Angelopoulos, A., E.~Cand{\`e}s, and R.~J. Tibshirani (2024).
\newblock Conformal pid control for time series prediction.
\newblock In {\em Adv. Neural Inf. Process. Syst. 2024}, Volume~36.

\bibitem[\protect\citeauthoryear{Barber, Cand{\`e}s, Ramdas, and
  Tibshirani}{Barber et~al.}{2021}]{foygel2021limits}
Barber, R.~F., E.~J. Cand{\`e}s, A.~Ramdas, and R.~J. Tibshirani (2021).
\newblock The limits of distribution-free conditional predictive inference.
\newblock {\em Inf. Inference\/}~{\em 10\/}(2), 455--482.

\bibitem[\protect\citeauthoryear{Barber, Cand{\`e}s, Ramdas, and
  Tibshirani}{Barber et~al.}{2023}]{barber2022conformal}
Barber, R.~F., E.~J. Cand{\`e}s, A.~Ramdas, and R.~J. Tibshirani (2023).
\newblock Conformal prediction beyond exchangeability.
\newblock {\em Ann. Stat.\/}~{\em 51\/}(2), 816--845.

\bibitem[\protect\citeauthoryear{Bates, Cand{\`e}s, Lei, Romano, and
  Sesia}{Bates et~al.}{2023}]{bates2023testing}
Bates, S., E.~Cand{\`e}s, L.~Lei, Y.~Romano, and M.~Sesia (2023).
\newblock Testing for outliers with conformal p-values.
\newblock {\em Ann. Stat.\/}~{\em 51\/}(1), 149--178.

\bibitem[\protect\citeauthoryear{Bhattacharjee and M{\"u}ller}{Bhattacharjee
  and M{\"u}ller}{2023}]{bhattacharjee2023single}
Bhattacharjee, S. and H.-G. M{\"u}ller (2023).
\newblock Single index fr{\'e}chet regression.
\newblock {\em Ann. Stat.\/}~{\em 51\/}(4), 1770--1798.

\bibitem[\protect\citeauthoryear{Bigot, Gouet, Klein, and L{\'o}pez}{Bigot
  et~al.}{2017}]{bigo:17}
Bigot, J., R.~Gouet, T.~Klein, and A.~L{\'o}pez (2017).
\newblock {Geodesic {PCA} in the {W}asserstein space by convex {PCA}}.
\newblock {\em Ann. inst. Henri Poincare (B) Probab. Stat.\/}~{\em 53}, 1--26.

\bibitem[\protect\citeauthoryear{Billera, Holmes, and Vogtmann}{Billera
  et~al.}{2001}]{billera2001geometry}
Billera, L.~J., S.~P. Holmes, and K.~Vogtmann (2001).
\newblock Geometry of the space of phylogenetic trees.
\newblock {\em Adv. Appl. Math.\/}~{\em 27\/}(4), 733--767.

\bibitem[\protect\citeauthoryear{Cand{\`e}s, Lei, and Ren}{Cand{\`e}s
  et~al.}{2023}]{ConformalSurvivalAnalysis}
Cand{\`e}s, E., L.~Lei, and Z.~Ren (2023).
\newblock Conformalized survival analysis.
\newblock {\em J. R. Stat. Soc. Ser. B Stat. Methodol.\/}~{\em 85\/}(1),
  24--45.

\bibitem[\protect\citeauthoryear{Chang}{Chang}{1989}]{chang1989spherical}
Chang, T. (1989).
\newblock Spherical regression with errors in variables.
\newblock {\em Ann. Stat.\/}~{\em 17\/}(1), 293--306.

\bibitem[\protect\citeauthoryear{Chen, Hall, and M{\"u}ller}{Chen
  et~al.}{2011}]{chen2011single}
Chen, D., P.~Hall, and H.-G. M{\"u}ller (2011).
\newblock {Single and multiple index functional regression models with
  nonparametric link}.
\newblock {\em Ann. Stat.\/}~{\em 39\/}(3), 1720--1747.

\bibitem[\protect\citeauthoryear{Chen, Lin, and M{\"u}ller}{Chen
  et~al.}{2023}]{chen2021wasserstein}
Chen, Y., Z.~Lin, and H.-G. M{\"u}ller (2023).
\newblock Wasserstein regression.
\newblock {\em J. Amer. Statist. Assoc.\/}~{\em 118\/}(542), 869--882.

\bibitem[\protect\citeauthoryear{Chernozhukov, W{\"u}thrich, and
  Zhu}{Chernozhukov et~al.}{2021}]{chernozhukov2021distributional}
Chernozhukov, V., K.~W{\"u}thrich, and Y.~Zhu (2021).
\newblock Distributional conformal prediction.
\newblock {\em Proc. Natl. Acad. Sci. U. S. A.\/}~{\em 118\/}(48), e2107794118.

\bibitem[\protect\citeauthoryear{Choi and Hall}{Choi and
  Hall}{1998}]{choi1998bias}
Choi, E. and P.~Hall (1998).
\newblock On bias reduction in local linear smoothing.
\newblock {\em Biometrika\/}~{\em 85\/}(2), 333--345.

\bibitem[\protect\citeauthoryear{Cornea, Zhu, Kim, and Ibrahim}{Cornea
  et~al.}{2017}]{cornea2017regression}
Cornea, E., H.~Zhu, P.~Kim, and J.~G. Ibrahim (2017).
\newblock Regression models on riemannian symmetric spaces.
\newblock {\em J. R. Stat. Soc. Ser. B Stat. Methodol.\/}~{\em 79\/}(2),
  463--482.

\bibitem[\protect\citeauthoryear{Du, Guo, Sun, and Zou}{Du
  et~al.}{2023}]{du2020}
Du, L., X.~Guo, W.~Sun, and C.~Zou (2023).
\newblock False discovery rate control under general dependence by symmetrized
  data aggregation.
\newblock {\em J. Amer. Statist. Assoc.\/}~{\em 118\/}(541), 607--621.

\bibitem[\protect\citeauthoryear{Dubey, Chen, and M{\"u}ller}{Dubey
  et~al.}{2024}]{dubey2022depth}
Dubey, P., Y.~Chen, and H.-G. M{\"u}ller (2024).
\newblock Metric statistics: Exploration and inference for random objects with
  distance profiles.
\newblock {\em Ann. Stat.\/}~{\em 52\/}(2), 757--792.

\bibitem[\protect\citeauthoryear{Dubey and M{\"u}ller}{Dubey and
  M{\"u}ller}{2020}]{dubey2020functional}
Dubey, P. and H.-G. M{\"u}ller (2020).
\newblock Functional models for time-varying random objects.
\newblock {\em J. R. Stat. Soc. Ser. B Stat. Methodol.\/}~{\em 82\/}(2),
  275--327.

\bibitem[\protect\citeauthoryear{Fan}{Fan}{1993}]{fan1993local}
Fan, J. (1993).
\newblock Local linear regression smoothers and their minimax efficiencies.
\newblock {\em Ann. Stat.\/}~{\em 21\/}(1), 196--216.

\bibitem[\protect\citeauthoryear{Fan and Gijbels}{Fan and
  Gijbels}{1992}]{fan1992variable}
Fan, J. and I.~Gijbels (1992).
\newblock Variable bandwidth and local linear regression smoothers.
\newblock {\em Ann. Stat.\/}~{\em 20\/}(4), 2008--2036.

\bibitem[\protect\citeauthoryear{Ferraty, Park, and Vieu}{Ferraty
  et~al.}{2011}]{ferraty2011estimation}
Ferraty, F., J.~Park, and P.~Vieu (2011).
\newblock Estimation of a functional single index model.
\newblock In {\em Recent Advances in Functional Data Analysis and Related
  Topics}, pp.\  111--116. Springer.

\bibitem[\protect\citeauthoryear{Fisher, Lewis, and Embleton}{Fisher
  et~al.}{1993}]{fisher1993statistical}
Fisher, N.~I., T.~Lewis, and B.~J. Embleton (1993).
\newblock {\em Statistical Analysis of Spherical Data}.
\newblock Cambridge University Press.

\bibitem[\protect\citeauthoryear{Fletcher}{Fletcher}{2013}]{thomas2013geodesic}
Fletcher, T.~P. (2013).
\newblock Geodesic regression and the theory of least squares on riemannian
  manifolds.
\newblock {\em Int. J. Comput. Vis.\/}~{\em 105}, 171--185.

\bibitem[\protect\citeauthoryear{Gao and Wellner}{Gao and
  Wellner}{2009}]{gao2009rate}
Gao, F. and J.~A. Wellner (2009).
\newblock On the rate of convergence of the maximum likelihood estimator of
  ak-monotone density.
\newblock {\em Sci. China Math.\/}~{\em 52\/}(7), 1525--1538.

\bibitem[\protect\citeauthoryear{Gibbs and Cand{\`e}s}{Gibbs and
  Cand{\`e}s}{2021}]{gibbs2021adaptive}
Gibbs, I. and E.~Cand{\`e}s (2021).
\newblock Adaptive conformal inference under distribution shift.
\newblock In {\em Adv. Neural Inf. Process. Syst. 2021}, pp.\  1660--1672.

\bibitem[\protect\citeauthoryear{Gibbs and Cand{\`e}s}{Gibbs and
  Cand{\`e}s}{2024}]{gibbs2022conformal}
Gibbs, I. and E.~J. Cand{\`e}s (2024).
\newblock Conformal inference for online prediction with arbitrary distribution
  shifts.
\newblock {\em J. Mach. Learn. Res.\/}~{\em 25\/}(162), 1--36.

\bibitem[\protect\citeauthoryear{Hall}{Hall}{1989}]{hall1989projection}
Hall, P. (1989).
\newblock On projection pursuit regression.
\newblock {\em Ann. Stat.\/}~{\em 17\/}(2), 573--588.

\bibitem[\protect\citeauthoryear{Hall and Marron}{Hall and
  Marron}{1997}]{hall1997role}
Hall, P. and J.~S. Marron (1997).
\newblock On the role of the shrinkage parameter in local linear smoothing.
\newblock {\em Probab. Theory Relat. Fields\/}~{\em 108}, 495--516.

\bibitem[\protect\citeauthoryear{Hall, Wolff, and Yao}{Hall
  et~al.}{1999}]{hall1999methods}
Hall, P., R.~C. Wolff, and Q.~Yao (1999).
\newblock Methods for estimating a conditional distribution function.
\newblock {\em J. Amer. Statist. Assoc.\/}~{\em 94\/}(445), 154--163.

\bibitem[\protect\citeauthoryear{Hu and Lei}{Hu and Lei}{2024}]{hu2023two}
Hu, X. and J.~Lei (2024).
\newblock A two-sample conditional distribution test using conformal prediction
  and weighted rank sum.
\newblock {\em J. Amer. Statist. Assoc.\/}~{\em 119\/}(546), 1136--1154.

\bibitem[\protect\citeauthoryear{Ichimura}{Ichimura}{1993}]{ichimura1993semiparametric}
Ichimura, H. (1993).
\newblock Semiparametric least squares (sls) and weighted sls estimation of
  single-index models.
\newblock {\em J. Econom.\/}~{\em 58\/}(1-2), 71--120.

\bibitem[\protect\citeauthoryear{Izbicki, Shimizu, and Stern}{Izbicki
  et~al.}{2022}]{izbicki2022cd}
Izbicki, R., G.~Shimizu, and R.~B. Stern (2022).
\newblock Cd-split and hpd-split: Efficient conformal regions in high
  dimensions.
\newblock {\em J. Mach. Learn. Res.\/}~{\em 23\/}(87), 1--32.

\bibitem[\protect\citeauthoryear{Jiang and Wang}{Jiang and
  Wang}{2011}]{jiang2011functional}
Jiang, C.-R. and J.-L. Wang (2011).
\newblock {Functional single index models for longitudinal data}.
\newblock {\em Ann. Stat.\/}~{\em 39\/}(1), 362--388.

\bibitem[\protect\citeauthoryear{Kuchibhotla and Patra}{Kuchibhotla and
  Patra}{2020}]{kuchibhotla2020efficient}
Kuchibhotla, A.~K. and R.~K. Patra (2020).
\newblock Efficient estimation in single index models through smoothing
  splines.
\newblock {\em Bernoulli\/}~{\em 26\/}(2), 1587--1618.

\bibitem[\protect\citeauthoryear{Lei, G'Sell, Rinaldo, Tibshirani, and
  Wasserman}{Lei et~al.}{2018}]{lei2018distribution}
Lei, J., M.~G'Sell, A.~Rinaldo, R.~J. Tibshirani, and L.~Wasserman (2018).
\newblock Distribution-free predictive inference for regression.
\newblock {\em J. Amer. Statist. Assoc.\/}~{\em 113\/}(523), 1094--1111.

\bibitem[\protect\citeauthoryear{Lei, Robins, and Wasserman}{Lei
  et~al.}{2013}]{lei2013distribution}
Lei, J., J.~Robins, and L.~Wasserman (2013).
\newblock Distribution-free prediction sets.
\newblock {\em J. Amer. Statist. Assoc.\/}~{\em 108\/}(501), 278--287.

\bibitem[\protect\citeauthoryear{Lei and Wasserman}{Lei and
  Wasserman}{2014}]{lei2014distribution}
Lei, J. and L.~Wasserman (2014).
\newblock Distribution-free prediction bands for non-parametric regression.
\newblock {\em J. R. Stat. Soc. Ser. B Stat. Methodol.\/}~{\em 76\/}(1),
  71--96.

\bibitem[\protect\citeauthoryear{Lin and M{\"u}ller}{Lin and
  M{\"u}ller}{2021}]{lin2021total}
Lin, Z. and H.-G. M{\"u}ller (2021).
\newblock Total variation regularized fr{\'e}chet regression for metric-space
  valued data.
\newblock {\em Ann. Stat.\/}~{\em 49\/}(6), 3510--3533.

\bibitem[\protect\citeauthoryear{Meinshausen, Meier, and
  B{\"u}hlmann}{Meinshausen et~al.}{2009}]{meinshausen2009}
Meinshausen, N., L.~Meier, and P.~B{\"u}hlmann (2009).
\newblock P-values for high-dimensional regression.
\newblock {\em J. Amer. Statist. Assoc.\/}~{\em 104\/}(488), 1671--1681.

\bibitem[\protect\citeauthoryear{Petersen and M{\"u}ller}{Petersen and
  M{\"u}ller}{2016}]{petersen2016functional}
Petersen, A. and H.-G. M{\"u}ller (2016).
\newblock Functional data analysis for density functions by transformation to a
  hilbert space.
\newblock {\em Ann. Stat.\/}~{\em 44\/}(1), 183--218.

\bibitem[\protect\citeauthoryear{Petersen and M{\"u}ller}{Petersen and
  M{\"u}ller}{2019}]{petersen2019frechet}
Petersen, A. and H.-G. M{\"u}ller (2019).
\newblock Fr{\'e}chet regression for random objects with euclidean predictors.
\newblock {\em Ann. Stat.\/}~{\em 47\/}(2), 691--719.

\bibitem[\protect\citeauthoryear{Petersen, Zhang, and Kokoszka}{Petersen
  et~al.}{2022}]{pete:22}
Petersen, A., C.~Zhang, and P.~Kokoszka (2022).
\newblock {Modeling probability density functions as data objects}.
\newblock {\em Econom. Stat.\/}~{\em 21}, 159--178.

\bibitem[\protect\citeauthoryear{Romano, Patterson, and Cand{\`e}s}{Romano
  et~al.}{2019}]{romano2019conformalized}
Romano, Y., E.~Patterson, and E.~Cand{\`e}s (2019).
\newblock Conformalized quantile regression.
\newblock In {\em Adv. Neural Inf. Process. Syst. 2019}, Volume~32.

\bibitem[\protect\citeauthoryear{Santambrogio}{Santambrogio}{2015}]{santambrogio2015optimal}
Santambrogio, F. (2015).
\newblock Optimal transport for applied mathematicians.
\newblock {\em Birk{\"a}user, NY\/}~{\em 55\/}(58-63), 94.

\bibitem[\protect\citeauthoryear{Sesia and Cand{\`e}s}{Sesia and
  Cand{\`e}s}{2020}]{sesia2020comparison}
Sesia, M. and E.~J. Cand{\`e}s (2020).
\newblock A comparison of some conformal quantile regression methods.
\newblock {\em Stat\/}~{\em 9\/}(1), e261.

\bibitem[\protect\citeauthoryear{Villani et~al.}{Villani
  et~al.}{2009}]{villani2009optimal}
Villani, C. et~al. (2009).
\newblock {\em Optimal Transport: Old and New}.
\newblock Springer.

\bibitem[\protect\citeauthoryear{Vovk}{Vovk}{2012}]{vovk2012conditional}
Vovk, V. (2012).
\newblock Conditional validity of inductive conformal predictors.
\newblock In {\em Asian Conf. Mach. Learn.}, pp.\  475--490.

\bibitem[\protect\citeauthoryear{Vovk}{Vovk}{2021}]{vovk2021conformal}
Vovk, V. (2021, 08--10 Sep).
\newblock Conformal testing in a binary model situation.
\newblock In {\em Proceedings of the Tenth Symposium on Conformal and
  Probabilistic Prediction and Applications}, Volume 152 of {\em Proc. Mach.
  Learn. Res.}, pp.\  131--150.

\bibitem[\protect\citeauthoryear{Vovk, Gammerman, and Shafer}{Vovk
  et~al.}{2005}]{vovk2005algorithmic}
Vovk, V., A.~Gammerman, and G.~Shafer (2005).
\newblock {\em Algorithmic Learning in a Random World}, Volume~29.
\newblock Springer.

\bibitem[\protect\citeauthoryear{Vovk, Nouretdinov, and Gammerman}{Vovk
  et~al.}{2009}]{vovk2009line}
Vovk, V., I.~Nouretdinov, and A.~Gammerman (2009).
\newblock On-line predictive linear regression.
\newblock {\em Ann. Stat.\/}, 1566--1590.

\bibitem[\protect\citeauthoryear{Vovk, Petej, Nouretdinov, Ahlberg, Carlsson,
  and Gammerman}{Vovk et~al.}{2021}]{vovk2021retrain}
Vovk, V., I.~Petej, I.~Nouretdinov, E.~Ahlberg, L.~Carlsson, and A.~Gammerman
  (2021).
\newblock Retrain or not retrain: Conformal test martingales for change-point
  detection.
\newblock In {\em Conformal and Probabilistic Prediction and Applications},
  pp.\  191--210.

\bibitem[\protect\citeauthoryear{Wang, Zhu, Pan, Zhu, and Zhang}{Wang
  et~al.}{2024}]{wang:23:3}
Wang, X., J.~Zhu, W.~Pan, J.~Zhu, and H.~Zhang (2024).
\newblock Nonparametric statistical inference via metric distribution function
  in metric spaces.
\newblock {\em J. Amer. Statist. Assoc.\/}~{\em 119\/}(548), 2772--2784.

\bibitem[\protect\citeauthoryear{Wasserman and Roeder}{Wasserman and
  Roeder}{2009}]{wasserman2009}
Wasserman, L. and K.~Roeder (2009).
\newblock High dimensional variable selection.
\newblock {\em Ann. Stat.\/}~{\em 37\/}(5A), 2178--2201.

\bibitem[\protect\citeauthoryear{Yang, Cand{\`e}s, and Lei}{Yang
  et~al.}{2024}]{yang2024bellman}
Yang, Z., E.~Cand{\`e}s, and L.~Lei (2024).
\newblock Bellman conformal inference: calibrating prediction intervals for
  time series.
\newblock {\em arXiv preprint arXiv:2402.05203\/}.

\bibitem[\protect\citeauthoryear{Yuan, Zhu, Lin, and Marron}{Yuan
  et~al.}{2012}]{yuan2012local}
Yuan, Y., H.~Zhu, W.~Lin, and J.~S. Marron (2012).
\newblock Local polynomial regression for symmetric positive definite matrices.
\newblock {\em J. R. Stat. Soc. Ser. B Stat. Methodol.\/}~{\em 74\/}(4),
  697--719.

\bibitem[\protect\citeauthoryear{Zhou and He}{Zhou and
  He}{2008}]{zhou2008dimension}
Zhou, J. and X.~He (2008).
\newblock {Dimension reduction based on constrained canonical correlation and
  variable filtering}.
\newblock {\em Ann. Stat.\/}~{\em 36\/}(4), 1649--1668.

\bibitem[\protect\citeauthoryear{Zhu and M{\"u}ller}{Zhu and
  M{\"u}ller}{2023}]{zhu2021autoregressive}
Zhu, C. and H.-G. M{\"u}ller (2023).
\newblock Autoregressive optimal transport models.
\newblock {\em J. R. Stat. Soc. Ser. B Stat. Methodol.\/}~{\em 85\/}(3),
  1012--1033.

\bibitem[\protect\citeauthoryear{Zhu and Zhu}{Zhu and
  Zhu}{2009}]{zhu2009distribution}
Zhu, L.-P. and L.-X. Zhu (2009).
\newblock On distribution-weighted partial least squares with diverging number
  of highly correlated predictors.
\newblock {\em J. R. Stat. Soc. Ser. B Stat. Methodol.\/}~{\em 71\/}(2),
  525--548.

\bibitem[\protect\citeauthoryear{Zou, Wang, and Li}{Zou et~al.}{2020}]{zou2020}
Zou, C., G.~Wang, and R.~Li (2020).
\newblock Consistent selection of the number of change-points via
  sample-splitting.
\newblock {\em Ann. Stat.\/}~{\em 48\/}(1), 413--439.

\end{thebibliography}

\end{document}